\documentclass[12pt]{article}
\usepackage[top=1in, bottom=1in, left=1in, right=1in]{geometry}
\usepackage[english]{babel}
\usepackage{atbegshi}
\usepackage{amsmath,amssymb,amsbsy,amstext, amsthm, simplewick}
\usepackage{graphicx}
\usepackage{amsfonts}
\usepackage[small]{caption}
\usepackage{upgreek}
\usepackage[titletoc]{appendix}
\usepackage{setspace}
\usepackage{comment}
\usepackage{tensor}
\usepackage[usenames,dvipsnames,table]{xcolor}
\usepackage[colorlinks=true,urlcolor=blue
,anchorcolor=blue,citecolor=blue,filecolor=blue,linkcolor=blue,menucolor=blue
]{hyperref}

\usepackage{parskip}

\def\pa{\partial}

\newcommand{\newc}{\newcommand}
\newc{\un}[1]{\underline{#1}}
\newc{\fpi}{f_{\pi}}
\newc{\etap}{\eta^{\prime}}
\newc{\llll}{\langle\lambda\lambda\rangle}
\newc{\FFd}{F^a\tilde F^a}
\newc{\qbar}{{\overline q}}
\newc{\TR}{{\rm Tr}}
\newc{\Kahler}{K\"ahler }
\newc{\Zbb}{{\mathbb Z}}
\newc{\Rt}{{{\mathbb R}^3}}
\newc{\Rf}{{{\mathbb R}^4}}
\newc{\Sth}{{{\mathbb S}^3}}
\newc{\SthSo}{{{\mathbb S}^3\times{\mathbb S}^1}}
\newc{\Stw}{{{\mathbb S}^2}}
\newc{\StwSo}{{{\mathbb S}^2\times{\mathbb S}^1}}
\newc{\So}{{{\mathbb S}^1}}
\newc{\zt}{{{\mathbb Z}_2}}
\newc{\RtSo}{{{\mathbb R}^3\times{\mathbb S}^1}}
\newc{\RfSo}{{{\mathbb R}^4\times{\mathbb S}^1}}
\newc{\scriminus}{{\cal I}^-}
\newc{\scriplus}{{\cal I}^+}
\newc{\mpl}{M_p}
\newc{\Ricci}{\mathcal{R}}
\newc{\bv}{\phi}
\newc{\calU}{{\cal U}}
\newc{\calK}{K}
\newc{\calUi}{{\cal U}^{-1}}
\newc{\calG}{{\cal G}}
\newc{\calH}{{\cal H}}
\newc{\calI}{{\cal I}}
\newc{\calL}{{\cal L}}
\newc{\calO}{{\cal O}}
\newc{\calQ}{{\cal Q}}
\newc{\calOb}{{\cal O}^\dagger}
\newc{\hphi}{{\hat\phi}}
\newc{\bes}[1]{\begin{equation} \begin{split} #1 \end{split} \end{equation}}

\theoremstyle{plain}
\theoremstyle{plain} 
\theoremstyle{plain} 
\theoremstyle{plain}
\theoremstyle{plain}
\theoremstyle{plain}

\newc{\pd}[1]{{\color{Red}#1}}
\newc{\hz}[1]{{\color{Green}#1}}

\renewcommand{\title}[1]{{\Large\bf\flushleft{#1}}\vspace*{3ex}\\}
\renewcommand{\author}[2]{{\noindent\hspace*{2.5em}\large#1}
                     \footnote{Electronic mail: $\mathtt{#2}$}\\}

\newcommand{\beq}{\begin{equation}}
\newcommand{\eeq}{\end{equation}}

\begin{document}
\begin{titlepage}

\vskip 2.2cm

\begin{center}

{\Large \bf Quantum Properties of Non-Dirichlet Boundary Conditions in Gravity}
\vskip 1.2cm

{Patrick Draper\footnote{pdraper@illinois.edu}$^{,a}$, Manthos Karydas \footnote{mkarydas@berkeley.edu}$^{,b}$, and Hao Zhang\footnote{haoz17@illinois.edu}$^{,a}$ }
~\\
\vskip 1cm
%{$^{(a)}$ 
{$^{a}$Illinois Center for Advanced Studies of the Universe \& Department of Physics,\\ University of Illinois, 1110 West Green St., Urbana IL 61801, U.S.A.}\\
\vspace{0.3cm}
{$^{b}$Physics Division, Lawrence Berkeley National Laboratory, Berkeley, CA 94720}\\
\vskip 4pt
%\tableofcontents

\vskip 1.5cm

\begin{abstract}
The Euclidean path integral for gravity is enriched by the addition of boundaries, which provide useful probes of thermodynamic properties. Common boundary conditions include Dirichlet conditions on the boundary induced metric; microcanonical conditions, which refers to fixing some components of the Brown-York boundary stress tensor; and conformal conditions, in which the conformal structure of the induced metric and the trace of the extrinsic curvature are fixed. Boundaries  also present interesting problems of consistency. The Dirichlet problem is known, under various (and generally different) conditions, to be inconsistent with perturbative quantization of graviton fluctuations; to exhibit thermodynamic instability; or to require infinite fine-tuning in the presence of matter fluctuations. We extend some of these results to other boundary conditions.
We find that similarly to the Dirichlet problem, the graviton fluctuation operator is not elliptic with microcanonical boundaries, and the non-elliptic modes correspond to ``boundary-moving diffeomorphisms." However, we argue that  microcanonical factorization of path integrals -- essentially, the insertion of microcanonical constraints on two-sided surfaces in the bulk -- is not affected by the same issues of ellipticity. We also show that  for a variety of matter field boundary conditions, matter fluctuations renormalize the gravitational bulk and boundary terms differently, so that the classical microcanonical or conformal variational problems are not preserved unless an infinite fine-tuning is performed. 
\end{abstract}

\end{center}

\vskip 1.0 cm

\end{titlepage}

\setcounter{footnote}{0} 
\setcounter{page}{1}
\setcounter{section}{0} \setcounter{subsection}{0}
\setcounter{subsubsection}{0}
\setcounter{figure}{0}

\section{Introduction}

Semiclassical treatment of the Euclidean path integral provides a powerful probe of thermodynamic and quantum aspects of gravity. The introduction of boundaries, which may be one-sided or two-sided,\footnote{A two-sided boundary may be thought of as a constraint in the path integral imposed on data on a bulk co-dimension one surface.} provides an additional probe of the theory. 

At the leading order in $\hbar$, the role of boundary terms in the action is to control the classical variational problem. Various classes of useful boundary conditions are known, including the Dirichlet/canonical boundary condition, the microcanonical boundary condition, and the conformal condition. Of these, one most commonly encounters  the Dirichlet problem. In cases with a time-like boundary Killing vector, the Dirichlet problem may be thought of as fixing the boundary temperature, and has been enormously useful in studying the thermodynamic properties of horizons \cite{PhysRevD.33.2092,PhysRevD.15.2752}. The microcanonical boundary condition introduced by Brown and York \cite{PhysRevD.47.1420} is thought of as the Legendre transform of the Dirichlet case, and when the Dirichlet problem corresponds to a canonical ensemble interpretation of the path integral, the microcanonical problem corresponds to the microcanonical ensemble. In some cases the microcanonical problem maybe well-defined even when the Dirichlet problem is not. For example, gravity in a cavity at positive cosmological constant has a lowest action solution of negative heat capacity, indicating an inconsistency of the canonical ensemble. Meanwhile the microcanonical problem still makes sense \cite{Draper_2022}.

At the one loop order, interesting new effects of boundary conditions arise.  One needs to impose boundary conditions on the fluctuations such that the fluctuation operator is suitably well-behaved. For example, with some sets of boundary conditions the fluctuation operator may fail to be elliptic, which in many cases indicates that it will not admit an inverse or a determinant~\cite{Avramidi:1997sh,Anderson_2008,Witten:2018lgb}. In the case of graviton fluctuations subject to   Dirichlet boundary conditions, the Laplace operator is not elliptic, which has led to renewed interest in the conformal boundary condition for which the Laplacian is elliptic~\cite{Anderson_2008,Witten:2018lgb}. In~\cite{Liu:2024ymn} a family of generalized conformal boundary terms $I=I_{EH}-\frac{\Theta}{8\pi G}\int_{\partial M}d^{D-1}x\sqrt{\gamma}K$ were introduced, where $\Theta=\frac{2p(D-1)-1}{(D-1)(2p-1)}$, and $p$ is a parameter. 
The ordinary conformal boundary term and the GHY boundary term are recovered when $\Theta$ approaches $(D-1)^{-1}$ and one, respectively, and for general $D$, the generalized boundary condition is elliptic when $p$ is finite~\cite{Liu:2024ymn}.

Matter fluctuations also impact the gravitational boundary conditions.   At one loop both  bulk and boundary terms in the gravitational action acquire ultraviolet sensitivity.  Curiously, the UV sensitivity may not preserve the set of well-posed variational problems associated with the bare action, leading to a ``renormalization of the boundary conditions." It was shown in Refs.~\cite{Barvinsky_1996,Jacobson_2014} that the Dirichlet problem renormalizes to itself when integrating out minimally coupled scalars or the Maxwell field with heat kernel methods, but for  non-minimally coupled matter with Dirichlet boundary condition, the relation between bulk and boundary divergences is different from the usual factor of $2$ between the Einstein-Hilbert term and the GHY term.  More recently Ref.~\cite{Neri_2023}, investigated how matter fluctuations renormalize higher-derivative operators in the effective action. 

Here we extend some of these results to other  boundary conditions. In section 2 we study ellipticity of the graviton fluctuation operator with microcanonical boundaries. We show that again it fails to be strongly elliptic, and the failure of ellipticity is associated with short wavelength fluctuations that correspond physically to ``boundary moving diffs." However, in the case of two-sided boundaries, corresponding to constraints in the path integral, or to factorization of the path integral with a microcanonical ``set of states," we argue that the failure of ellipticity is no longer a problem. because boundary moving diffs for a two-sided boundary are in fact real gauge transformations. 

In section 3 we study matter fluctuations. We show that with minimally coupled scalar matter obeying a general class of ``oblique," Robin-type boundary conditions, the bulk terms in the gravitational action and the boundary terms associated with the microcanonical condition  generically exhibit different sensitivities to ultraviolet renormalization effects. This indicates that microcanonical gravitational boundaries have a naturalness problem. Like all naturalness problems, it can be ignored, or addressed in technical applications by fine-tuning, but the level of fine tuning is in a sense infinite, since only one value of the boundary/bulk coupling ratio is consistent with a given variational problem. The issue is also not just about  power-law cutoff sensitivity. Although the leading
divergences are power-law, integrating out a massive particle produces similar logarithmic and finite corrections
 dependent on the particle mass. 

If the scalar has non-minimal coupling to the Ricci scalar we find that standard heat kernel results cannot be applied, because the boundary terms still depend on the scalar field after the bulk Lagrangian is written in the form $-\phi^*\nabla^2\phi$. However, in the non-minimal coupling case the gravitational conformal boundary condition can be imposed instead, and the effective action can be derived with the heat kernel method. We find that the conformal boundary condition-violating corrections can cancel out, but only under a rather bizarre set of conditions. In most cases the conformal boundary condition is not preserved in a Wilsonian effective action without fine-tuning, which raises a puzzle for the status of strong ellipticity. Finally, we study the Maxwell field, and show that the naturalness issue arises both for microcanonical and conformal gravitational boundaries.

We have included two Appendices that supplement the material of the main text. In Appendix~\ref{App:Micro} we consider the microcanonical gravitational variational problem for minimally coupled complex scalar in subsection~\ref{App:A}, and non-minimally coupled complex scalar in subsection~\ref{App:B}. In Appendix~\ref{App:C} we consider the variational problem with conformal gravitational boundary conditions for a non-minimally coupled complex scalar.

\section{Non-Ellipticity of Microcanical Boundary Conditions}
\label{sec:Non-Ellipticity of Microcanical Boundary Conditions}
In this section we consider the ellipticity of the fluctuation operator in pure Euclidean gravity with microcanonical boundary conditions.\footnote{Our conventions in this section are as follows. We work in $D$-dimensional Euclidean spacetime, with $D$-dimensional indices $MN...$ To define microcanonical boundary conditions we also need a ``time" foliation, corresponding to a time function $t$, and it is convenient to use the ADM decomposition for this foliation. We denote the  slices of fixed $t$ by $\Sigma_t$ and the boundary of each timeslice by $\mathcal{B}$, which is codimension-2 in the bulk spacetime. $(D-1)$-dimensional indices $\mu\nu$ are used for spatial tensors and $(D-2)$-dimensional indices $ab$ are used for tensors on $\mathcal{B}$. Later, in Sec.~3, $(D-1)$-dimensional indices $ij$ will be used for coordinates on the codimension-1 boundary of the Euclidean spacetime.} Previously it was shown in~\cite{Anderson_2008,Witten:2018lgb} that Dirichlet boundary conditions are not elliptic, and in~\cite{Witten:2018lgb} that the conformal boundary condition is elliptic. 
We use the method described in~\cite{Witten:2018lgb}. 

The bulk Euclidean Einstein-Hilbert  action is 
\beq
\label{EH_action_bulk}
S_{\text{EH}}=-\frac{1}{2\kappa}\int_{{\cal M}} d^Dx \sqrt{g}(R-2\Lambda)\,,
\eeq
where $\kappa=8\pi G_{N}$.
We expand the metric around a classical solution $\Bar{g}_{MN}$,
\begin{align}
    \label{gaugetransform}
g_{MN}=\Bar{g}_{MN}+h_{MN}.
\end{align}
To second order in the fluctuation $h_{MN}$, the EH action \eqref{EH_action_bulk} becomes (see e.g. \cite{bastianelli2013oneloop}) 
\begin{align*}
    S_{\text{EH}}=&-\frac{1}{2\kappa}\int_{\cal M} d^Dx\sqrt{\Bar{g}}\bigg\{\frac{1}{4}h^{MN }(\Bar{\nabla}^2+2\Lambda)h_{MN}-\frac{1}{8}h(\Bar{\nabla}^2+2\Lambda)h+\frac{1}{2}(\Bar{\nabla}^M h_{NM}-\frac{1}{2}\Bar{\nabla}_M h)^2\\
    +&\frac{1}{2}h^{ML}h^{NS}\Bar{R}_{MNLS}+\frac{1}{2}(h^{ML}h\indices{_{L}^{N}}-hh^{MN})\Bar{R}_{MN}+\frac{1}{8}(h^2-2h^{MN}h_{MN})\Bar{R}\bigg\}.
\end{align*}
Here we dropped boundary terms, $\Bar{\nabla}$ and barred quantities are evaluated with respect to a background metric that satisfies the classical equations of motion $\bar R_{MN}= \frac{2\Lambda}{D-2}\bar g_{MN}$.\footnote{Indices are raised using the background metric, e.g. $h\indices{^{M}_{N}}:= \bar g^{M K}h_{K N}$ and $h= \bar g_{MN}h^{MN}$.} In the harmonic gauge
\beq
\label{harmonic_gauge_operator}
T_{M}(h):=\Bar{\nabla}^N h_{MN}-\frac{1}{2}\partial_M h=0\,,
\eeq
the equation of motion is
\begin{align}
    \label{gauged EOM}
    \Bar{\nabla}^2 h_{MN} + 2\bar R_{ML N K}h^{L K}=0.
\end{align}
Ellipticity of an operator in the presence of a boundary is a condition that holds pointwise, and it can be assessed by examining the properties of the zero modes of the highest derivative terms in the operator. For Einstein gravity in the harmonic gauge only the $\bar g^{MN}\partial_M\partial_N$ term in the fluctuation operator \eqref{gauged EOM} is needed. Furthermore we may focus on points near the boundary in coordinates where the metric is locally the ordinary Euclidean flat metric and the boundary is located at $x_\perp=0$~\cite{Witten:2018lgb}. This can be thought of as ``zooming in" on an arbitrary point near the boundary where the manifold is locally the half space ${\mathbb{R}}^{D}_{+}$. In this section, ``boundary" specifically refers to the ``timelike" portion of the boundary where the microcanonical boundary conditions are to be imposed. The relevant zero modes of the Laplace operator \eqref{gauged EOM}, if any, can be written in the form 
\begin{align}
\label{elliptic_type_zero_modes}
h_{MN}=\alpha_{MN}\exp[i\Vec{k}\cdot\Vec{x}-\vert \vec{k}\vert x_\perp],
\end{align}
for some $(D-1)$-dimensional vector $\vec{k}$ and constant matrix $a_{MN}$.\footnote{$\vec{k}\cdot \vec{x}$ is defined using the Euclidean flat metric.} 
The question is whether there are any such fluctuations that \textbf{(a)} satisfy the boundary conditions, \textbf{(b)} satisfy the harmonic gauge, and \textbf{(c)} are not pure gauge. If there are, then there is at least a pileup of infinitely many near-zero modes that may make it impossible to define perturbation theory (for example, the spectral decomposition of the Green function may not converge.)

To answer the question posed above we first need to define the microcanonical boundary conditions. It is convenient to work in the ADM decomposition~\cite{Arnowitt:1962hi} of the metric adapted to a ``time" foliation $t:{\cal M}\to \mathbb{R}$ 
\begin{align}
\label{metric}
    ds^2=g_{MN}dx^M dx^N=N^2dt^2+\gamma_{\mu\nu}(dx^\mu+V^\nu dt)(dx^\mu+V^\nu dt),
\end{align}
where $N$ is the lapse function, $V^\mu$ is the shift vector, and $\gamma_{\mu\nu}$ is the induced metric on each timeslice. The microcanonical boundary conditions proposed by Brown and York~\cite{PhysRevD.47.1420,PhysRevD.47.1407} are\footnote{The notation $A\,|$ means the quantity $A$ is evaluated on the boundary.}
\begin{align}
    \label{microcanonical conditions}
\delta(\varepsilon\sqrt{s})\,|=\delta(J_a\sqrt{s})\,|=\delta s_{ab}\,|=0,\quad\left(\text{microcanonical BCs}\right)\,, 
\end{align}
where the energy surface density $\varepsilon$, momentum surface density $J^{a}$ and $s_{ab}$ are
\beq
\begin{split}
\varepsilon &= \frac{1}{\kappa}k\quad,\quad
J_{a}= \frac{2}{\sqrt{\gamma}}s_{ab}n_{\lambda}P^{b\lambda}\quad,\quad s_{ab}=\gamma_{ab}\,.
\end{split}
\eeq
Here $s_{ab}$, $n_\mu$, and $k$ are the induced metric, unit normal vector, and the trace of the 
extrinsic curvature of the codimension-2 boundary $\mathcal{B}$ respectively, $P^{\mu\nu}$ is the conjugate ADM momentum of $\gamma_{\mu\nu}$. Their explicit formulas can be found in~\cite{PhysRevD.47.1420},
\beq
\label{k_P_formulas}
\begin{split}
k&=\frac{1}{\sqrt{\gamma}}\pa_{\mu}\left(\sqrt{\gamma}n^{\mu}\right)\,,\,P^{\mu\nu}=- \frac{1}{4\kappa}\frac{\sqrt{\gamma}}{N}(\gamma^{\mu\nu}\gamma^{\kappa\lambda} - \gamma^{\mu\kappa}\gamma^{\nu\lambda})(\dot \gamma_{\kappa\lambda}- 2D_{(\kappa}N_{\lambda)})\,,
\end{split}
\eeq
where $\dot \gamma_{\mu\nu}:=\pa_{t}\gamma_{\mu\nu}$ and $D_{\mu}$ is the induced covariant derivative on $\Sigma_{t}$.

In arriving at these quantities, it is assumed that the boundary is orthogonal to the foliation, which amounts to 
\beq
\label{orthogonality_condition}
n_{\mu}V^{\mu}\,|=0,\quad\left(\text{orthogonality condition}\right)\,.
\eeq
We will see this condition can always be restored (if it holds for the background) by an infinitesimal pure gauge transformation, even in the presence fluctuations of the form \eqref{elliptic_type_zero_modes}. So in our context we can assume \eqref{orthogonality_condition} without any loss of generality.

From the ADM decomposition~\eqref{metric} we can express the ADM variables $(N, N^{\mu},\gamma_{\mu\nu})$ in terms of $g_{MN}$ and thus find an expression for their variation in terms of the fluctuations $h_{MN}$ and background values. Henceforth we use $N$, $V^\mu$, and $\gamma_{\mu\nu}$ to denote the background lapse, shift, and induced metric, and we denote their (off-shell) fluctuations by $\delta N$, $\delta V^\mu$, $\delta \gamma_{\mu\nu}$. We find
\beq
\begin{split}
    \label{variationofLandV}
    &\delta N=\frac{1}{2N}(h_{tt}-2V^\sigma(h_{\sigma t}-h_{\sigma\nu}V^\nu)-V^\mu h_{\mu\nu}V^\nu)\,,\\
    &\delta V^\mu=\gamma^{\mu\sigma}(h_{\sigma t}-h_{\sigma\nu}V^\nu)\,,\\
    &\delta \gamma_{\mu\nu}= h_{\mu\nu}\,.
\end{split}
\eeq

In the locally flat coordinates near the boundary we parameterize $x^{\mu}=(x^{a},t,x_{\perp})$ with $a=1,\dots,D-2$. The boundary is located at $x_{\perp}=0$ and $(x^{a},t)$ are now the boundary coordinates. Since the metric is locally flat we can set 
\beq
V^{\mu}=0~,~ N=1~,~\gamma\indices{^{\mu}_{\nu}}=\delta\indices{^{\mu}_{\nu}} \,,
\eeq
for the background shift, lapse and induced metric. The background unit normal vector is $n_{\mu}=(n_{a},n_{t},n_{\perp})=(0,0,n_{\perp})$. Then the formulas in \eqref{variationofLandV} simplify to
\beq
    \label{variaofNV}
    \delta N=\frac{1}{2}h_{tt}\quad,\quad \delta V^\mu=h\indices{^{\mu}_{t}}\quad,\quad \delta\gamma_{\mu\nu}=h_{\mu\nu}.
\eeq
We have all the ingredients to compute what the microcanonical boundary conditions and orthogonality conditions imply for the metric fluctuations. Making use of Eq.~\eqref{variaofNV} and Eq.~\eqref{k_P_formulas} we can compute Eqs.~\eqref{microcanonical conditions} and the variation of Eq.~\eqref{orthogonality_condition}. We find
\begin{alignat}{2}
\label{Micro_boundary_conditions_fluctuations_1}
\left(\frac{1}{2}\sum_{a=1}^{D-2}\pa_{\perp}h_{aa} -  \sum_{a=1}^{D-2}\pa_{a}(h_{a\perp})\right)\,|&=0\qquad && \left(\delta\varepsilon\,|=0\right)\,,\\
\label{Micro_boundary_conditions_fluctuations_2}
(\pa_{t} h_{a\perp}- \pa_{a}h_{t\perp}-\pa_{\perp}h_{t a})\,|&=0\qquad&& \left(\delta J_a\,|=0\right)\,,\\
\label{Micro_boundary_conditions_fluctuations_3}
h_{ab}\,|&=0\quad&& \left(\delta s_{ab}\,|=0\right)\,,\\
\label{Micro_boundary_conditions_fluctuations_4}
h_{t\perp}\,|&=0\qquad &&\left(\text{orthogonality condition}\right)\,.
\end{alignat}
The above equations should be satisfied by the fluctuations if they are to preserve the microcanonical boundary conditions including orthogonality. If we also impose the harmonic gauge condition $T_{M}(h)=0$ to the fluctuations we find from Eq.~\eqref{harmonic_gauge_operator} the following conditions
\begin{alignat}{2}
\label{Micro_boundary_conditions_fluctuations_5}
\frac{1}{2}\pa_{t}h_{tt}+ \pa_{\perp}h_{t\perp} + \sum_{a=1}^{D-2}\pa_{a}h_{t a}  - \frac{1}{2}\left(\pa_{t}h_{\perp\perp}+ \sum_{a=1}^{D-2}\pa_{t}h_{aa}\right)&=0\quad&& \left(T_{t}(h)=0\right)\,,\\
\label{Micro_boundary_conditions_fluctuations_6}
\pa_{t}h_{\perp t} + \frac{1}{2}\pa_{\perp}h_{\perp\perp} + \sum_{a=1}^{D-2}\pa_{a}h_{\perp a} - \frac{1}{2}\left(\pa_{\perp}h_{tt}+ \sum_{a=1}^{D-2}\pa_{\perp}h_{aa} \right)&=0\quad &&\left(T_{\perp}(h)=0\right)\,,\\
\label{Micro_boundary_conditions_fluctuations_7}
\pa_{t}h_{at}+ \pa_{\perp}h_{a\perp}+ \sum_{b=1}^{D-2}\pa_{b}h_{b a} -\frac{1}{2}\left(\pa_{a}h_{tt}+ \pa_{a}h_{\perp\perp}+ \sum_{b=1}^{D-2}\pa_{a}h_{bb}\right)&=0\quad &&\left(T_{a}(h)=0\right)\,.
\end{alignat}
As mentioned earlier, the relevant zero modes of the Laplace operator for non-ellipticity take the form $h_{MN}=\alpha_{MN}\exp[ik_{t}t+ i\sum\limits{_{a=1}^{D-2}}k_{a}x^{a} -\vert \vec{k}\vert x_\perp]$ with $\vec k = (k_{t}, k^{a})$. If we substitute this form into Eqs.~\eqref{Micro_boundary_conditions_fluctuations_1}-\eqref{Micro_boundary_conditions_fluctuations_7} we find the following set of conditions for the constants $\alpha_{MN}$ 
\begin{alignat}{2}
\label{Micro_boundary_conditions_fluctuations_1_waves}
\frac{1}{2}\sum_{a=1}^{D-2}|\vec k|\alpha_{aa}+i\sum_{a=1}^{D-2}k^{a}\alpha_{a\perp} &=0\quad && \left(\delta\varepsilon\,|=0\right)\,,\\
\label{Micro_boundary_conditions_fluctuations_2_waves}
ik_{t} \alpha_{\perp a}- i k^{a}\alpha_{t\perp}+ |\vec k|\alpha_{t a}&=0\quad&& \left(\delta J_a\,|=0\right)\,,\\
\label{Micro_boundary_conditions_fluctuations_3_waves}
\alpha_{ab}&=0\quad&& \left(\delta s_{ab}\,|=0\right)\,,\\
\label{Micro_boundary_conditions_fluctuations_4_waves}
\alpha_{t\perp}&=0\quad &&\left(\text{orthogonality}\right)\,,\\
\label{Micro_boundary_conditions_fluctuations_5_waves}
\frac{1}{2}ik_{t}\alpha_{tt}- |\vec k|\alpha_{t\perp}+ i\sum_{a=1}^{D-2}k^{a}\alpha_{t a} -\frac{1}{2}i k_{t}\alpha_{\perp\perp} - \frac{1}{2}ik_{t}\sum_{a=1}^{D-2}\alpha_{aa}&=0\quad&& \left(T_{t}(h)=0\right)\,,\\
\label{Micro_boundary_conditions_fluctuations_6_waves}
ik_{t}\alpha_{\perp t}-\frac{1}{2}|\vec k |\alpha_{\perp\perp} + i \sum_{a=1}^{D-2}k^{a}\alpha_{\perp a}+ \frac{1}{2} |\vec k |\alpha_{tt}+\frac{1}{2}|\vec k|\sum_{a=1}^{D-2}\alpha_{aa}&=0\quad&& \left(T_{\perp}(h)=0\right)\,,\\
\label{Micro_boundary_conditions_fluctuations_7_waves}
ik_{t}\alpha_{at}-|\vec k|\alpha_{a\perp}+ i \sum_{b=1}^{D-2}k^{b}\alpha_{ba} - \frac{1}{2}ik^{a}\alpha_{tt} -\frac{1}{2}ik^{a}\alpha_{\perp\perp} -\frac{1}{2}i k^{a} \sum_{b=1}^{D-2}\alpha_{bb} &=0\quad &&\left(T_{a}(h)=0\right)\,.
\end{alignat}
The above equations have the following non-zero general solution
\beq
\label{non_elliptic_modes_general_solution_microcanonical_alphas}
\alpha_{\perp a}= \beta_{a},\quad\alpha_{ta}= - i \frac{k_{t}}{|\vec k|}\beta_{a}\,,\quad
\alpha_{tt}=\alpha_{\perp\perp}=\alpha\,, 
\eeq
with $|\vec k|= |k_{t}|$. The corresponding metric fluctuations are 
\beq
\label{non_elliptic_modes_general_solution_microcanonical}
\begin{split}
h_{\perp a}= \beta_{a}e^{ik_{t}t - |\vec k| x_{\perp}}~,&~ h_{t a}= - i \frac{k_{t}}{|\vec k|}\beta_{a}e^{ik_{t}t - |\vec k| x_{\perp}}\,,\\
h_{tt}=h_{\perp\perp}&=\alpha e^{i k_{t}t - |\vec k| x_{\perp}}\,.
\end{split}
\eeq
We note that there is a continuous family of zero modes parameterized by $\beta_{a}$ with $a=1,\dots,D-2$ and $\alpha$. To establish the failure of ellipticity we need to show that within the set of solutions given in Eq.~\eqref{non_elliptic_modes_general_solution_microcanonical} there exists at least one non-zero solution that is not included in the set of proper gauge transformations. It is not hard to check that the fluctuations in Eq.~\eqref{non_elliptic_modes_general_solution_microcanonical} can be generated by an infinitesimal diffeomorphism $x^{M}\to x^{M}+ \xi^{M}$ with
\beq
\begin{split}
\label{diffeo_of_non_elliptic_mode_micro}
\xi_{t}= \frac{\alpha}{2 i k_{t}} e^{i k_{t}t - |\vec k| x_{\perp}}~&,~\xi_{\perp}=- \frac{\alpha}{2|\vec k|}e^{i k_{t}t - |\vec k| x_{\perp}}\,,\\
\xi_{a}=-\frac{1}{|\vec k|}\beta_{a}& e^{i k_{t}t - |\vec k| x_{\perp}}\,.
\end{split}
\eeq
For $\alpha=0, \beta_{a}\neq0$ the above diffeomorphism is a proper gauge transformation and the corresponding fluctuations in \eqref{non_elliptic_modes_general_solution_microcanonical}. However, for $\alpha\neq 0, \beta_{a}=0$ the diffeomorphism is ``boundary-moving diff", not included in the set of proper gauge transformations. The corresponding fluctuations satisfy all three requirements $\textbf{(a)}-\textbf{(b)}-\textbf{(c)}$ mentioned earlier which proves that the microcanonical boundary conditions are not elliptic. We conclude that  the failure of ellipticity is due to arbitrarily short-distance fluctuations of the metric near the boundary that would be pure gauge (generated by proper diffs) if it were not for the presence of a physical boundary. We will return to this point again below when we discuss factorization and unphysical boundaries.

The ``boundary moving diff" interpretation also holds for the non-elliptic zero modes of the canonical  (Dirichlet) boundary condition. The non-elliptic modes for the Dirichlet boundary conditions take the form \cite{Witten:2018lgb}
\beq
h_{MN}=\alpha_{MN}e^{i\vec k\cdot \vec x - |\vec k|x_{\perp}}\,,
\eeq
with the only non zero components $\alpha_{\perp \perp}=\beta$ and $\alpha_{\mu\perp}=-\frac{i}{2|\vec k|}k_{\mu}\beta$. It is straightforward to check that the following boundary moving diff
\begin{align}
    \xi_\perp=\frac{1}{2|\vec k|}\beta e^{{i\Vec{k}\cdot\Vec{x}-\vert k\vert x^\perp}},
\end{align}
can generate the above fluctuation.  We expect the boundary moving diff interpretation to be a general feature for any boundary condition, namely, all non elliptic modes, if they exist, are generated by boundary moving diffeomorphisms.

Let us return to a technical detail glossed over in the microcanonical anlalysis above. In the derivation (and in using the Brown-York definitions of the surface energy and momentum flux) we assumed that the boundary is orthogonal to the time foliation with respect to which the energy is defined.  This condition is not preserved by a fluctuation of the form given in Eq.~\eqref{elliptic_type_zero_modes}, but we can show that it can be restored by an appropriate boundary-preserving diffeomorphism. 
Consider the following diff:
\begin{align}
    \label{diffappA}
    \xi_t=\frac{\alpha_{t\perp}}{\vert \Vec{k}\vert}e^{ik_tt+i\sum\limits{_{a=1}^{D-2}}k^ax^a-\vert\Vec{k}\vert x_\perp}.
\end{align}
The corresponding metric fluctutations are 
\begin{align*}
    &h_{t\perp}=-\alpha_{t\perp}e^{ik_tt+i\sum\limits{_{a=1}^{D-2}}k^ax^a-\vert\Vec{k}\vert x_\perp},\\
    &h_{t a}=\frac{ik^a}{\vert\Vec{k}\vert}\alpha_{t\perp}e^{ik_tt+i\sum\limits{_{a=1}^{D-2}}k^bx^b-\vert\Vec{k}\vert x_\perp},\\
    &h_{tt}=\frac{ik_t}{\vert\Vec{k}\vert}\alpha_{t\perp}e^{ik_tt+i\sum\limits{_{a=1}^{D-2}}k^ax^a-\vert\Vec{k}\vert x_\perp}.\\
\end{align*}
It is easy to check above fluctuations satisfy the microcanonical boundary  conditions and the harmonic gauge. Since it does not move the boundary the above diffeomorphism is a pure gauge transformation. For a non-orthogonal fluctuation one has $\alpha_{t \perp}\neq 0$ (see Eq.~\eqref{Micro_boundary_conditions_fluctuations_4_waves}), and the diffeo in Eq.~\eqref{diffappA} can be utilized to transform $\alpha_{t\perp}$ back to zero, thus restoring orthogonality.

\begin{figure}[!ht]
 \centering \includegraphics[width=0.8\textwidth]{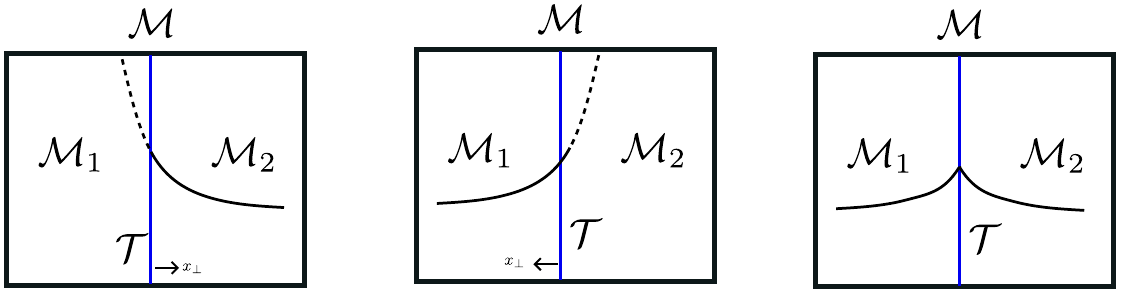}
  \caption{The manifold ${\cal M}$ and factorization surface ${\cal T}$ (blue). Left: Analytically extending (dotted line) the non-elliptic mode from ${\cal M}_{2}$ to ${\cal M}_{1}$. Middle: Analytically extending (dotted line) the non-elliptic mode from ${\cal M}_{1}$ to ${\cal M}_{2}$. Right: Pontwise gluing of the non-elliptic modes across ${\cal T}$. The mode does not satisfy Laplace equation on ${\cal T}$ due to the non-analyticity at $x_{\perp}=0$.}
  \label{fig:facto_non_elliptic}
\end{figure}

We conclude this section with some comments on boundary conditions and factorization surfaces. A factorization surface\footnote{We exclude corner cases where ${\cal T}\cap {\pa {\cal M}}\neq 0$.} is any cod-1 surface that partitions ${\cal M}$ into two regions ${\cal M}_{1}$, ${\cal M}_{2}$ with ${\cal M}= {\cal M}_{1}\cup {\cal M}_{2}$ such that the Euclidean path integral factorizes 
\beq
Z= \int_{{\cal T}}D{\cal B}_{i}Z({\cal M}_{1}|{\cal B}_{i})Z({\cal M}_{2}|{\cal B}_{i})\,.
\eeq
Here the integrand is written as a product of two path integrals over ${\cal M}_{1,2}$ with fixed boundary data ${\cal B}_i$ at the interface ${\cal T}$.
It turns out \cite{Draper:2022xzl,Draper:2023bhg} that the fields belonging to ${\cal B}_{i}$ are those that need to be fixed in order for the action to have a well-defined variational problem. Each elliptic mode on either side of ${\cal T}$ can be continued to the other side of ${\cal T}$ by extending the domain to include $x_{\perp}\le 0$, see Figure \ref{fig:facto_non_elliptic}. In this section we have shown that these non-elliptic modes when ${\cal B}_{i}$ are either  Dirichlet or microcanonical boundary conditions are generated by boundary moving diffs. For ${\cal T}$ these are diffeomorphisms that ``move" the factorization surface ${\cal T}$ and correspond to actual gauge transformations. This is in contrast to the case where the surface is an actual boundary. Moreover, if the modes on either side of ${\cal T}$ are glued together pointwise on ${\cal T}$, as shown in Figure \ref{fig:facto_non_elliptic}, they do not satisfy the Laplace equation on ${\cal T}$. We thus conclude that the non-elliptic modes on either side of ${\cal T}$ are pure gauge and thus both Dirichlet and microcanonical boundary conditions ${\cal B}_{i}$ can be used to factorize the path integral.

\section{Renormalization of microcanonical and conformal boundary conditions}
\label{sec:Renormalization of microcanonical and conformal boundary conditions}

We turn now to quantum fluctuations of matter coupled to gravity in the presence of boundaries. 
Ultraviolet fluctuations of matter fields renormalize the local operators in the gravitational effective action, including boundary operators. The leading terms in the derivative expansion have been calculated for a large variety of matter fields and boundary conditions, as tabulated and reviewed in~\cite{Vassilevich_2003}.

It was noted in~\cite{Jacobson_2014,Barvinsky_1996,Becker_2012}  that short-distance fluctuations can disturb the balance between bulk and boundary terms, altering the form of the classical boundary value problem at different renormalization scales. With minimally coupled free fields, however, the gravitational Dirichlet problem is preserved under renormalization. Here we will study the same question in the case of microcanonical and conformal boundary conditions.

\subsection{Review of heat kernel method}
\label{sec:Review of Heat Kernel method}
We start by reviewing the heat kernel method for computing effective actions. We present the relevant formulas used in subsequent sections and refer to \cite{Vassilevich_2003} for more details. 

Consider a Riemannian manifold ${\cal M}$ of dimension $D$ and boundary $\pa {\cal M}$, along with a complex scalar field $\phi : {\cal M}\to \mathbb{C}$. We will treat $\phi(x)$ as a section of a rank-2 real vector bundle V over ${\cal M}$, namely for $\phi(x)=\alpha(x)+ i\beta(x)$ we denote
\bes{
\phi_{A}(x)= \begin{pmatrix}\alpha\\ \beta\end{pmatrix}~,~~A=1,2\,.
}
For $\phi_{1},\phi_{2}$ we define the inner product
\bes{
\label{inner_product}
\langle \phi_{1},\phi_{2} \rangle= \int_{{\cal M}}d^{D}x\sqrt{g}\,\TR\,  \phi_{1}\phi_{2}\,,
}
where the trace is defined as $\TR\,\phi_{1}\phi_{2}\equiv \delta_{AB}\phi_{1,A}\phi_{2,B}$, and $\delta_{AB}$ the 2-dimensional Euclidean metric. 

Consider a Laplace type operator $\cal D$ that is symmetric with respect to the inner product \eqref{inner_product}, namely it satisfies $\langle \phi_{1},{\cal D}\phi_{2} \rangle= \langle {\cal D}\phi_{1},\phi_{2} \rangle$. Introducing back the components the operator ${\cal D}$ will take the general form:
\bes{
\label{cal_D_operator}
{\cal D}=-\left[\delta_{AB}\left(g^{MN}\nabla_{M}\nabla_{N} - m^{2}\right) + E_{AB}(g)\right] 
}

Consider now an action for the metric + matter that is quadratic in the complex scalar field:
\bes{
I[g,\phi]=I_{\rm grav}(g)+ \int d^{D}x \sqrt{g}\TR\, \phi {\cal D}\phi + I_{\pa}(g,\phi)\,.
}
The boundary term is needed so that there is a well-defined variational problem for both metric and scalar field variations. We can write the boundary conditions formally as
\bes{
\label{bc's_general}
{\cal C}_{1}[g]|_{\pa{\cal M}}&=0~~ \textbf{(I)},~~
{\cal C}_{2}[g,\phi]|_{\pa{\cal M}}=0~~\textbf{(II)}\,.
}
The form of the boundary conditions above is not the most general one could write; we have assumed $C_{1}[g]|_{\pa {\cal M}}$ is independent of the scalar field $\phi$. This ``decoupling" of the gravitational from the scalar boundary conditions will be essential for applying the heat kernel method. We will encounter this issue in the microcanonical case for both minimal and non-minimal cases (Sections~\ref{subsec:Formalism and the example of scalar field}, \ref{subsec:Non-minimally coupled scalar + microcanonical boundary}).

The full path integral is
\bes{
Z_{\rm total}= \int [D g][D\phi]e^{- I[g,\phi]}\,.
}
The question we want to address is whether the boundary condition \textbf{(I)} for the metric can be preserved after integrating out the scalar field with the boundary condition \textbf{(II)} of \eqref{bc's_general}.
For this we substitute $\phi\to \phi_{0}+ \phi$ into the action, expanding it up to quadratic order in the fluctuations, and then perform the Gaussian path integral over the scalar field fluctuations. We assume that $\phi_0$ satisfies the scalar field equations of motion for arbitrary metric ${\cal D}\phi_{0}=0$. 

Note that expanding \textbf{(II)} to linear order in the scalar field imply the boundary conditions for the scalar field fluctuations, which in our context take the form ``oblique" Robin-like boundary conditions~\cite{DMMcAvity_1991,Vassilevich_2003} (see also \cite{Dowker:1997mn}):\footnote{In this section  Latin lowercase letters $i,j,\dots$ refer to coordinates on the codimension-1 boundary.}
\bes{
\label{eq:oblique_bc_heat_kernel}
[\nabla_n\delta_{AB} +\frac{1}{2}(\Gamma^i_{AB} D_i+D_i\Gamma^i_{AB})+ S_{AB}]\phi_{B} \,\big|_{\pa{\cal M}}=0\,.
}
Here $\Gamma^i$ and $S$ may be functions of the boundary coordinates, $\nabla_{n}:=n^{M}\nabla_{M}$ is the  outward normal derivative to the boundary\footnote{Note this differs from the conventions of \cite{Vassilevich_2003} where the unit normal $n$ is taken to be inward pointing.} and $D_{i}$ is the induced covariant derivative (computed from the induced metric). 
It is easy to check that the condition of ${\cal D}$ being symmetric implies\footnote{We assume no corner terms appear when integrating by parts the boundary derivative $D_{i}$.}
\bes{
\label{symmetric_condition_real_rep}
S_{AB}=S_{BA}~~,~~ \Gamma^{i}_{AB}=-\Gamma^{i}_{BA}\,.
}
Integrating out the scalar field fluctuations we obtain
\bes{
Z_{\rm total}= \int [D g_{MN}] e^{- I_{\rm grav}[g]} e^{- S_{\rm eff}[g]} \,,
}
where the effective action $S_{\rm eff}[g]$ is
\bes{
\label{eq:effective_integral_eigenvalues}
S_{\rm eff}[g]&=\frac{1}{2}\ln(\text{det}\,{\cal D})\\&
=-\frac{1}{2}\sum_{n}\int_{0}^{\infty}\frac{dt}{t}e^{-t \lambda_{n}}\,,
}
where $\lambda_n$ denote the eigenvalues of ${\cal D}$ (see Eq.~\eqref{D_eigenvalue_equation} below). It is important to note that the symmetric property $\langle \phi_{1},{\cal D}\phi_{2} \rangle= \langle {\cal D}\phi_{1},\phi_{2} \rangle$ is necessary for validity of the determinant formula \eqref{eq:effective_integral_eigenvalues}.

The heat kernel $K_{AB}(t;x,y)$ is defined by the heat equation
\bes{
\label{eq:heat_kernel}
(\pa_{t}\delta_{AC} + {\cal D}_{x,AC})K_{CB}(t;x,y)=0\,,
}
with the initial condition $K_{AB}(0;x,y)= \delta_{AB}\delta (x-y)$ and the delta function definition $\int d^{D}y \sqrt{g} f(y)\delta (x-y)=f(x)$.

Assume now $\phi_{A,n}(x)$ are the eigenfunctions of ${\cal D}$ with the boundary conditions \eqref{eq:oblique_bc_heat_kernel}, satisfying
\bes{
\label{D_eigenvalue_equation}
{\cal D}_{AB}\phi_{B,n}(x)=\lambda_{n}\phi_{A,n}(x)\,,~~\lambda_{n}\equiv m^{2}+\tilde\lambda_{n}\,.
}
The set $\phi_{n}(x)$ forms a complete orthonormal set under the inner product \eqref{inner_product}, namely $\langle \phi_{n},\phi_{m}\rangle =\delta_{nm}$. The heat kernel can be expressed in terms of $\{\phi_{A,n}(x)$, $\lambda_{n}\}$ as
\bes{
\label{heat_kernel_eigenfunctions}
K_{AB}(t;x,y)= \sum_{n}e^{- t\lambda_{n}}\phi_{A,n}(x)\phi_{B,n}(y)\,,
}
which can be checked by direct substitution into \eqref{eq:heat_kernel}\,, and the initial condition is satisfied due to the completeness relation 
\bes{
\sum_{n}\phi_{A,n}(x)\phi_{B,n}(y)= \delta_{AB}\delta (x-y)\,.
}
From Eq.~\eqref{eq:effective_integral_eigenvalues} and Eq.~\eqref{heat_kernel_eigenfunctions} we can express $S_{\rm eff}[g]$ formally in terms of the heat kernel
\bes{
S_{\rm eff}[g]&= -\frac{1}{2}\int_{0}^{\infty}\frac{dt}{t}\int d^{D}x \sqrt{g}\left(\delta_{AB}K_{AB}(t;x,x)\right)\,.
}
The benefit of the above equation is that it can be expanded as an asymptotic series $t\to 0$ of local geometric invariants \cite{Vassilevich_2003}
\begin{align}
\label{Seff_heat_kernel}
S_{\rm eff} = -\frac{1}{2}\int^\infty_{\Lambda^{-2}}\frac{dt}{t}e^{-m^2t}\sum_{k\geq 0} a_k t^{(k-D)/2}\,,
\end{align}
where we introduced an ultraviolet cutoff $\Lambda$ on the lower bound of the integral to regularize the divergent expression. Assuming $[\Gamma^{i},\Gamma^{j}]=0$ the relevant coefficients in the asymptotic expansion of the heat kernel  are~\cite{DMMcAvity_1991,Vassilevich_2003}\footnote{Here we  restrict to the case of one scalar field and set a smoothing function $f$ equal to unity; more general expressions are given in~\cite{Vassilevich_2003}, but are not needed for our purposes.} \footnote{Note that the $\Gamma$'s we consider may not be covariantly constant, which is often a simplification assumed to reduce the number of invariants in the effective action: non-covariantly constant $\Gamma$ introduces terms containing $D\Gamma$. However, since those terms are of higher order in the $G_N$ expansion on dimensional grounds, we will not need to take them into consideration.}
\begin{align}
\label{0orderofHK_heat_kernel}
a_0&=1\,,\\
\label{1storderofHK_heat_kernel}
a_1&=\pm (4\pi)^{-(D-1)/2}\int_{\partial {\cal M}}d^{D-1}x\sqrt{\beta}\TR\left(\Tilde{\gamma}(\Gamma)\right),\\
\label{2ndorderofHK_heat_kernel}
a_2&=\frac{1}{6}(4\pi)^{-D/2}\bigg[\int_{{\cal M}}d^Dx\sqrt{g}\,\TR (6E+R)\nonumber\\&\quad
+\int_{\partial {\cal M}}d^{D-1}x\sqrt{\beta}\,\TR\left(b_0(\Gamma)K+b_2(\Gamma)S+\sigma(\Gamma)K_{ij}\Gamma^i\Gamma^j\right)\bigg]\,.
\end{align}
The $\pm$ sign in $a_1$ denotes oblique (+) or Dirichlet (-) boundary conditions.\footnote{For Dirichlet $S=0,\Gamma^i=0$ should be assumed in the formulas for $a_0$, $a_1$,$a_2$.} 
The coefficients above are functions of the magnitude of the matrix valued boundary condition vector $\Gamma^2=\Gamma^i\Gamma_{i}$, 
\begin{align}
\label{coefficientsofHK_heat_kernel}
\Tilde{\gamma}&=\frac{1}{4}\left[\frac{2}{\sqrt{1+\Gamma^2}}-1\right],\nonumber\\
    b_0&=6\left[\frac{1}{1+\Gamma^2}-\frac{1}{\sqrt{-\Gamma^2}}\text{arctanh}(\sqrt{-\Gamma^2})\right]+2\nonumber\,,\\
    b_2&=-\frac{12}{1+\Gamma^2}\nonumber\,,\\
    \sigma&=\frac{1}{\Gamma^2}(2-b_0)\,.
\end{align}
The extra minus sign in $b_{2}$ compared~\cite{Vassilevich_2003} is due to the fact that in our conventions the normal vector in Eq.~\eqref{eq:oblique_bc_heat_kernel} is outward pointing to the boundary.

\subsection{Minimally coupled scalar + microcanonical boundary}
\label{subsec:Formalism and the example of scalar field}
In this section, we investigate matter renormalization effects on the microcanonical boundary value problem. We consider a massless, minimally coupled complex scalar field with the  general class of ``oblique" Robin-like boundary conditions decribed in Section~\ref{sec:Review of Heat Kernel method} (see Eq.~\eqref{eq:oblique_bc_heat_kernel}),

\begin{align}
    [\nabla_n +\frac{1}{2}(\Gamma^i D_i+D_i\Gamma^i)+ S]\phi \,\big|_{\pa{\cal M}}=0.
    \label{GammaSbc}
\end{align}
Here $S$, $\Gamma^{i}$ are complex valued. To match with Eq.~\eqref{eq:oblique_bc_heat_kernel} we need to identify
\bes{
\label{S_Gamma_matrices_matching}
S_{AB}= \begin{pmatrix}
\text{Re }S&- \text{Im }S\\
\text{Im }S&~~\text{Re }S
\end{pmatrix}~,~\Gamma^{i}_{AB}= \begin{pmatrix}
\text{Re }\Gamma^i &- \text{Im }\Gamma^i\\
\text{Im }\Gamma^i&~~\text{Re }\Gamma^i
\end{pmatrix}\,.
}
The symmetric condition \eqref{symmetric_condition_real_rep} of ${\cal D}$ (see Eq.~\eqref{cal_D_operator}) implies $\text{Im }S=0$ and $\text{Re }\Gamma^{i}=0$. 
The oblique boundary conditions are elliptic\footnote{To see this, we can zoom in sufficiently into a local region of $\pa{\cal M}$ and check for solutions $\phi=e^{-|\vec k|x_{\perp}+ i \vec k\cdot \vec x}$ of Eq.~\eqref{GammaSbc}. Assuming $S^{*}=S$, $\Gamma^{i*}=-\Gamma^{i}$ we find a solution $|\vec k|=S/\left(1+ |\vec m|\cos\theta \right)$, with $m^{i}:= i \Gamma^{i}$ and $\theta$ the angle between $\vec m$ and $\vec k$. For $|\vec m|>1$ the solutions can have arbitrary large $|\vec{k}|$, violating ellipticity.} as long as $\Gamma:=|\Gamma_{i}\Gamma^{i}|<1$.  

We must first construct an appropriate action reflecting the boundary condition~(\ref{GammaSbc}) and a microcanonical condition for the gravitational fields.  The microcanonical action is constructed from the Dirichlet action by Legendre transform. We begin with a Dirichlet action\footnote{For simplicity we set $\Lambda=0$ in this section. It can be reinstated without affecting our main results.} for the gravitational fields and a general scalar boundary term:
\begin{align}
    \label{Euclideanaction}
    I=&\int_{\cal M} d^Dx\sqrt{g}\,\left[|\nabla\phi|^2+m^2|\phi|^2-\frac{1}{2\kappa}R \right]\nonumber\\
    +&\int_{\partial{\cal M}}d^{D-1}x\sqrt{\beta}\,[-\frac{1}{\kappa}K+a\phi^*\phi+
    (c^i\phi^*\partial_i\phi+\text{c.c.})].
\end{align}
Here $a$ and $c^i$ are boundary couplings that will determine the appropriate classical variational problem for $\phi$ (or vice versa), and $\beta$ is the codim-1 induced metric on $\partial {\cal M}$. $c^i$ is tangent to $\partial {\cal M}$, and we omit normal derivatives of $\phi$ to avoid conflict with the gravitational Dirichlet problem. Apart from possible corner terms, the variation of the scalar field yields a boundary variation (see Appendix~\ref{App:A})
\begin{align}
    \label{variationscalar}
    \delta I=&\int_{\partial M}d^{D-1}x\sqrt{\beta}\,[\delta\phi^*(n^M\partial_M\phi+a\phi+c^i\partial_i\phi-\frac{1}{\sqrt{\beta}}\partial_i(\sqrt{\beta}c^{i*}\phi))]\nonumber\\
    &+ \text{c.c.}+(\text{bulk term})\,,
\end{align}
where $n^M$ is the outwards pointing unit normal to the boundary. 

A relation between the boundary value problem, parametrized by $S$ and $\Gamma_i$ in Eq.~(\ref{GammaSbc}), and the boundary couplings $a$ and $c^i$, is now clear. We identify 
\begin{equation}
\label{relationbetweenrobandboundary}
\begin{split}
        S&=a\,,\\
    \Gamma^i&=2c^i\,,\\
    \text{Im }S&=\text{Im }a=0\,,\\
    \text{Re }\Gamma^i&=\text{Re }c^i=0\;.
\end{split}
\end{equation}
Eq.~(\ref{relationbetweenrobandboundary}) encodes the oblique boundary conditions for the canonical scalar bulk action $\int_M d^{D}x\sqrt{g} \left(|\pa_{M} \phi|^2+ m^2 |\phi|^2\right)$.

 Now we perform an on-shell variation of the action with respect to the lapse $N$ to determine the Brown-York energy. In Appendix~\ref{App:A} we provide more details of this procedure. 
Shortly we will see that it is useful to focus on cases where $c^i$ is normalized to have constant length and the shift vector is gauge-fixed to zero. In particular
we will be interested in the case where $c^i$ is proportional to the time vector $\un t= \pa/\pa t$ associated with the boundary foliation; namely we assume $\un c= i N^{-1}\un t$ from this point on. In that case, the scalar boundary couplings do depend on the metric, $c^i \propto 1/N$, where $N$ is the lapse function. 
Consequently there are other scalar contributions to $I_\text{micro}$ in addition to the $[k-K]$ structure. The Brown-York energy is 
\begin{align}
   -\int dt\int_{\partial\Sigma_t}d^{D-2}x\,\sqrt{s}\,\delta N[\frac{1}{\kappa}k-a\phi^*\phi]\equiv-\int dt\int_{\partial\Sigma_t}d^{D-2}x\sqrt{s}\,\delta N\,\varepsilon,
\label{byenergy}
\end{align}
so the microcanonical action in this case is\footnote{The ``(bulk)" term in Eq.~\eqref{Imicro2} refers to bulk integral together with boundary terms on the spacelike part of the boundary. See Appendix~\ref{App:A} for details.} 
\begin{align}
   I_\text{micro}=(\text{bulk})+ \int dt\int_{\partial\Sigma_t}d^{D-2}x\sqrt{\beta}\,\left(\frac{1}{\kappa}[k-K]+ \left(\phi^*c^i\partial_i\phi+\text{c.c.}\right)\right)\,.
   \label{Imicro2}
\end{align}
We emphasize again that in the derivation of Eq.~\eqref{Imicro2}, the shift was set to zero.

By construction, the microcanonical action is consistent with the scalar oblique boundary condition~(\ref{relationbetweenrobandboundary}),~(\ref{GammaSbc}) and fixed BY (Brown-York) energy~(\ref{byenergy}). However, there is one further consideration that leads us to take $a=S=0$. Ultimately we would like to integrate by parts the bulk kinetic term, $|\partial_M\phi|^2\rightarrow -\phi^*\nabla^2\phi $, so that we can integrate out $\phi$ fluctuations and evaluate a functional determinant using standard methods. This procedure only works if there is no residual $\phi$ boundary term after the integration by parts and also implies the microcanonical boundary conditions are ``decoupled", in the sense that the BY energy $\varepsilon$ should be independent of scalar field. We see immediately from Eq.~\eqref{byenergy} that this is the case only if $a=0$. In this case the BY energy reduces to its usual form in Einstein gravity: 
\begin{align}
    \varepsilon_{BY} = k/\kappa ~~~\text{for}~~~ S=a=0.
\end{align}

As usual, there is also some arbitrariness in $I_\text{micro}$ – we can add boundary terms that are functions of quantities that are fixed in the boundary value problem. However, this does not alter the  $[k-K]$ structure, which is the relevant piece for the question of renormalization of the boundary condition. 

The conclusion of this analysis is that we must attempt to choose an oblique scalar boundary condition, parametrized by $\Gamma^i$, such that the radiative corrections to the gravitational boundary terms have the form $(1/\kappa)\int dt \int_{\pa \Sigma_{t}}(k-K)$, with coefficient $1$ relative to the correction to the Einstein-Hilbert term $\sim -\frac{1}{2\kappa}\int_M R$.

The effective action $S_{\rm eff}$ is given in Eq.~\eqref{Seff_heat_kernel} and the relevant coefficients in Eqs.~\eqref{1storderofHK_heat_kernel}, \eqref{2ndorderofHK_heat_kernel}, where the boundary we consider here is the ``timelike" boundary\footnote{For the ``spacelike" part of the boundary one fixes the induced metric even in the microcanonical case. Dirichlet or Neumann boundary conditions for a minimally coupled scalar field suffice to preserve the gravitational boundary conditions for the spacelike part of the boundary \cite{Jacobson_2014}. One is allowed to do this because the boundary terms in the effective action are ``locally" determined by the boundary conditions of the scalar field. From  Eq.~\eqref{2ndorderofHK_heat_kernel} it is evident that the expressions are local boundary integrals involving the boundary functions $S(x),\Gamma^{i}(x)$.}. 
Because the coefficients $a_{1},a_{2}$ are complicated functions of $\Gamma^i$, in order to obtain a simple gravitational boundary value problem in the effective action, we must require that $\Gamma^2$ is  a constant.

To relate the microcanonical gravitational boundary terms in Eq.~(\ref{Imicro2}) to the boundary terms generated in the effective action~(\ref{2ndorderofHK_heat_kernel}), note that 
the $K-k$ structure can be rewritten as the $tt$ component of the extrinsic curvature of the  codimension-1 boundary~\cite{PhysRevD.47.1407,PhysRevD.47.1420}:
\begin{align}
    \label{calcK-k}
    K-k=-a^M n_M=u^M u^N K_{MN}\,,
\end{align}
where $u^M$ is the unit normal vector of the time slices, $a^N = u^M\nabla_M u^N$ $n_M$ is the unit normal to the boundary, and the shift is set to zero. 
So if we can choose oblique boundary conditions $\Gamma^i = i z u^i$ for constant real $z$, and then choose $z$ such that 
\begin{align}
b_0=0\,,
\end{align} 
then
the boundary term in the radiative correction Eq.~(\ref{2ndorderofHK_heat_kernel}) will have the correct form. 

Unfortunately,  there is no choice of  $z$ that satisfies these conditions: it is readily seen from Eq.~\eqref{coefficientsofHK_heat_kernel} that $b_0\geq 2$. A second issue is that the $a_1$ term in Eq.~\eqref{1storderofHK_heat_kernel} is also inconsistent with the microcanonical boundary terms. 

The issue is also not merely one of power-law ultraviolet divergences. Although the leading divergences are power-law, integrating out a massive particle produces logs and finite shifts of the same form. For example, in $D=4$, the $a_2$ term in Eq.~(\ref{Seff_heat_kernel}) provides
\begin{align}
-\frac{1}{2}a_2\int_{\Lambda^{-2}}^\infty\frac{dt}{t^2}e^{-m^2t}=-\frac{1}{2}a_2\left[\Lambda^2+m^2\log(\Lambda^2/m^2)+(\gamma-1)m^2+\calO(\Lambda^{-2})\right].
\end{align}

\subsection{Polemic Interlude}
What are we to make of this?  A conservative conclusion is  that with minimally coupled scalar matter, the microcanonical boundary condition can be applied either at the level of the bare action or the renormalized action, but not both. One can achieve a renormalized action consistent with any desired boundary condition by tuning separate counterterms for bulk and boundary curvature scalars. 
At a practical level, this does not appear to cause problems. 

However, it is quite disturbing if one wants to think of the gravitational effective action in a Wilsonian sense. To realize the microcanonical boundary condition in the renormalized action requires an infinite level of fine-tuning at the cutoff, including even contributions to the counterterms sensitive to infrared mass scales.  

We will also see that in simple cases the same holds for conformal boundary conditions. This has puzzling implications for the problem of ellipticity: whether the graviton fluctuation operator is strongly elliptic with conformal boundary conditions may depend on whether or not one has already integrated out massive matter degrees of freedom.

It is not clear what to make of all this, but it is tempting to speculate that ``radiative stability of boundary value problem" imposes some swampland-type constraints on effective field theories coupled to gravity on spacetimes with boundaries.

\subsection{Non-minimally coupled scalar + microcanonical boundary}
\label{subsec:Non-minimally coupled scalar + microcanonical boundary}

Now we turn to the case where the scalar field is non-minimally coupled to the gravitational field. 
The action of the non-minimally coupled scalar field with gravitational Dirichlet boundary conditions is \cite{Jacobson_2014}
\begin{align}
\label{micrononminimalcpling}
    I_{\rm n.m.}=&\int_M d^Dx\sqrt{g}\left(\vert\nabla\phi\vert^2+\phi^*(m^2-\xi R)\phi-\frac{1}{2\kappa}R\right)\nonumber\\
    &~~~~~+\frac{1}{\kappa}\int_{\partial M}d^{D-1}x\sqrt{\beta}\left(-K(2\kappa\xi|\phi|^2+1)+ a|\phi|^2 +\left(c^{i}\phi^*\partial_i\phi+\text{c.c.}\right)\right)\,.
\end{align}
The metric and scalar field boundary terms vanish given the following boundary conditions (see Appendix~\ref{App:B}) 
\bes{
\label{bcs_non_minimal}
\delta \beta_{ij}|_{\pa{\cal M}}=0~~,~~\nabla_{n}\phi + (a-2\xi K)\phi - \frac{1}{\sqrt{\beta}}\pa_{i}(\sqrt{\beta}c^{i*}\phi) + c^{i}\pa_{i}\phi \,|_{\pa {\cal M}}=0\,.
}
The boundary conditions are Dirichlet for the metric and of  oblique type for the scalar field. They match Eq. \eqref{GammaSbc} if we identify 
\beq
S=a-2\xi K~,~\Gamma^i=2c^i\,,
\eeq
with $\text{Im }S=\text{Re }\Gamma^i =\text{Re }c^i)=0$.

As in the minimally-coupled case we assume the shift vector vanishes and $\un c= i N^{-1}\un t$. The on-shell variation of the lapse function gives the quasilocal energy:
\begin{align}
&\int dt\int_{\pa\Sigma_{t}}d^{D-2}x\sqrt{s}\bigg[-\frac{1}{\kappa}k\left(1+ 2\kappa \xi |\phi|^{2}\right)+ a|\phi|^{2} -2\xi \nabla_{n}(|\phi|^2) \bigg]\delta N \\& \equiv- \int dt \int_{\pa \Sigma_{t}}d^{D-2}x\sqrt{s}\,\delta N \varepsilon_{\rm n.m}
\end{align}
with the non-minimal BY energy given by
\bes{
\label{BY_non_minimal}
\varepsilon_{\rm n.m}=\frac{1}{\kappa}k\left(1+ 2\kappa \xi |\phi|^{2}\right)- a|\phi|^{2} +2\xi \nabla_{n}(|\phi|^2)\,.
}
The microcanonical ensemble is now defined by fixing $\varepsilon_{\rm n.m.}$ instead of the lapse and keeping the same oblique boundary conditions for the scalar field as before.

In the non-minimal coupling case $\xi\neq 0$, even if $a=0$, we see that $\varepsilon_{\rm n.m.}$ depends on the scalar field, namely the boundary conditions are always ``coupled". This poses a serious obstruction to integrating the scalar field fluctuations with heat kernel methods because the boundary terms generated from integrating by parts the bulk terms $|\partial_M\phi|^2\rightarrow -\phi^*\nabla^2\phi $ do not cancel. In this case we are unable to investigate how the microcanonical boundary terms renormalize using standard heat kernel methods. It is possible that a suitable change of path integral variables might decouple the microcanonical boundary conditions and standard heat kernel methods could be applied. We leave such investigations for future work.

\subsection{Non-minimally coupled scalar + conformal boundary}
\label{subsec:Non-minimally coupled scalar + conformal boundary}
We start from an action of the form
\begin{align}
    \label{nmnmCB_v2}
     I_{\rm conf}=&\int_M d^Dx\sqrt{g}\left(\vert\nabla\phi\vert^2+\phi^*(m^2-\xi R)\phi-\frac{1}{2\kappa}R\right)\nonumber\\
    -&\frac{1}{\kappa}\int_{\partial M}d^{D-1}x\sqrt{\beta} \left(\frac{1}{D-1}K(1+2\kappa\xi\vert\phi\vert^2)+\kappa(aK+b)|\phi|^2\right)
\end{align}
where $a$ and $b$ are constants. The boundary terms in the variation are (see Appendix~\ref{App:C})
\bes{
\label{conformal_total_variation}
\delta I_{\rm conf}|_{\pa{\cal M}} &=
\int_{\pa {\cal M}}d^{D-1}x\sqrt{\beta}\left(-\xi\nabla_{n}|\phi|^2 - \frac{1}{2}b|\phi|^2\right)\beta^{ij} 
\delta\beta_{ij} \\&
+\int_{\pa {\cal M}}d^{D-1}x\sqrt{\beta}\bigg[\left(\frac{1}{2\kappa} +\xi |\phi|^2\right)(K^{ij}- \frac{1}{D-1}K \beta^{ij})- \frac{1}{2}a |\phi|^{2}K\beta^{ij} \bigg]\delta\beta_{ij}\\&
+\int_{\pa{\cal M}}d^{D-1}x |\phi|^{2}\left[2\xi\left(\frac{D-2}{D-1}\right)- a\right] \delta K \\&
+\int_{\pa{\cal M}}d^{D-1}x\sqrt{\beta}\left[\nabla_{n}\phi^{*} - b\phi^{*}- \frac{2}{D-1}K\xi\phi^{*}- aK \phi^{*}\right]\delta\phi \\&
+\int_{\pa{\cal M}}d^{D-1}x\sqrt{\beta}\left[\nabla_{n}\phi - b\phi - \frac{2}{D-1}K \xi \phi - aK \phi\right]\delta\phi^{*}\,.
}
Now let us see what requirements a conformal boundary condition for the metric imply for $a$ and $b$. It is sufficient to consider a general variation of the scalar and a conformal variation of the boundary metric, $\beta_{ij}\rightarrow(1+\sigma)\beta_{ij}$, and we require $\delta K=0$. We find
\bes{
\delta I_{\rm conf}|_{\pa{\cal M}} &= \int_{\pa {\cal M}}d^{D-1}x\sqrt{\beta}\left(2\xi\nabla_{n}|\phi|^2 + b|\phi|^2 + K a|\phi|^{2}\right)\left(\frac{1-D}{2}\right)\sigma\\&
+\int_{\pa{\cal M}}d^{D-1}x\sqrt{\beta}\left[\nabla_{n}\phi^{*} - b\phi^{*}- \frac{2}{D-1}K\xi\phi^{*}- aK \phi^{*}\right]\delta\phi \\&
+\int_{\pa{\cal M}}d^{D-1}x\sqrt{\beta}\left[\nabla_{n}\phi - b\phi - \frac{2}{D-1}K \xi \phi - aK \phi\right]\delta\phi^{*}\,.
}
Now let us impose a Robin-type boundary condition of the form
\begin{align}
\label{Robin_conformal_matter}
    \nabla_n\phi - (\psi K+\chi) \phi=0
\end{align}
where $\psi$ and $\chi$ are real constants. 
Then we may rearrange the boundary variation in the form
\bes{
\delta I_{\rm conf}|_{\pa{\cal M}} &= \int_{\pa {\cal M}}d^{D-1}x\sqrt{\beta}\left((4\xi\psi+ a)K + 4\xi \chi + b\right)\left(\frac{1-D}{2}\right)|\phi|^2\sigma\\&
+\int_{\pa{\cal M}}d^{D-1}x\sqrt{\beta}\left[\left(\psi- \frac{2}{D-1}\xi -a \right) K + \chi - b \right]\delta|\phi|^{2}\,.
}
Thus we find the relations
\bes{
4\xi\psi+ a &=0,~~
4\xi \chi + b=0\,,\\
\psi- \frac{2}{D-1}\xi -a &=0,~~
\chi - b=0\,,
}
with solutions
\bes{
a&=-\frac{8 \xi ^2}{(D-1) (4 \xi +1)},~~\psi=\frac{2 \xi }{(D-1) (4 \xi +1)},~~b=\chi=0\,,
}
for $\xi\neq-1/4$, and no solution for $\xi=-1/4$. With these choices, the action \eqref{nmnmCB_v2} has a well-defined  variational problem with $\delta K = \delta [\beta_{ij}]=0$, with $[\beta_{ij}]$ the conformal structure of the boundary metric and Robin-type boundary conditions
\bes{
\label{scalar_boundary_condition_conformal}
\nabla_{n}\phi - \frac{2\xi}{(D-1)(4\xi +1)}K\phi=0\,,
}
for the scalar field. Matching the above with Eq.~\eqref{eq:oblique_bc_heat_kernel} (see also Eqs. \eqref{GammaSbc} and \eqref{S_Gamma_matrices_matching}) we obtain
\bes{
S_{AB}=- \frac{2\xi}{(D-1)(4\xi +1)}K \delta_{AB}~,~\Gamma^{i}_{AB}=0\,.
}
For the action in Eq.~\eqref{nmnmCB_v2}, the operator ${\cal D}$ (see definition in Eq.~\eqref{cal_D_operator}) yields $E_{AB}= \xi R\, \delta_{AB}$.
From Eq.~\eqref{coefficientsofHK_heat_kernel} we thus find $b_0=2,\,b_2=-12$,\footnote{We have suppressed the identity matrix in the expressions for $b_0,b_{2}$.} and the coefficients of the effective action $S_{{\rm eff};\phi}$ in Eq.~\eqref{Seff_heat_kernel} are
\bes{
\label{heat_kernel_coef_phi_conformal}
a_{0\,;\phi}&=1\,,\\
a_{1\,;\phi}&=2(4\pi)^{-(D-1)/2}\int_{\pa{\cal M}}d^{D-1}x\sqrt{\beta}\,,\\
a_{2\,;\phi}&= \frac{1}{2}(4\pi)^{-D/2}\bigg[ \int_{{\cal M}}d^{D}x\sqrt{g}(6\xi  + 1)R + \int_{\pa{\cal M}}d^{D-1}x\sqrt{\beta}\left(2 + 12\frac{2\xi}{(D-1)(4\xi+1)}\right)K\bigg]\,.
}
The boundary term $a_{1;\phi}$ is problematic since it does not preserve the conformal boundary conditions $\delta[\beta_{ij}]=\delta K=0$. To cancel this term we need to introduce one other minimally coupled complex scalar field $\phi'$ of the same mass $m$ and with Dirichlet boundary condition $\phi'|_{\pa{\cal M}}=0$.\footnote{For this additional field we have the same matter action as Eq.~\eqref{nmnmCB_v2} with $\xi=a=b=0$. One can check that the gravitational conformal boundary conditions $\delta[\beta_{ij}]=\delta K=0$ plus Dirichlet for the complex scalar still yield a well defined variational problem.} This is because in (\ref{1storderofHK_heat_kernel}) the $a_1$ of the oblique boundary condition has different sign compared to the one with Dirichlet boundary condition. The coefficients of the effective action $S_{{\rm eff};\phi'}$ for the additional field $\phi'$ are 
\bes{
\label{heat_kernel_coef_phi_conformal}
a_{0\,;\phi'}&=1\,,\\
a_{1\,;\phi'}&=-2(4\pi)^{-(D-1)/2}\int_{\pa{\cal M}}d^{D-1}x\sqrt{\beta}\,,\\
a_{2\,;\phi'}&= \frac{1}{2}(4\pi)^{-D/2}\bigg[ \int_{{\cal M}}d^{D}x\sqrt{g}R + \int_{\pa{\cal M}}d^{D-1}x\sqrt{\beta}\left(2 K\right)\bigg]\,.
}
For the total effective action $S_{eff}= S_{eff;\phi}+S_{eff;\phi'}$ we see that $a_{1}\equiv a_{1;\phi}+ a_{1;\phi'}=0$ by construction. Requiring $a_{2}\equiv a_{2;\phi}+ a_{2;\phi'}$ to preserve the gravitational conformal boundary condition we obtain
 \begin{align}
 \label{Confomal_preserved_condtion}
     \frac{2+2+12\frac{2\xi}{(D-1)(1+4\xi)}}{1 +1+6\xi}=\frac{2}{D-1}\,,
 \end{align}
 where the numerator on the left hand side above contains the coefficients of the effective action proportional to the extrinsic curvature, and the denominator contains those proportional to the Ricci scalar. Solving Eq.~\eqref{Confomal_preserved_condtion} for the non-minimal coupling $\xi$ we find two possible solutions
 \begin{align}
 \label{solutions_xi_conformal}
     \xi_{\pm}=\frac{1}{24} \left[(4 D-5) \pm \sqrt{16 D^2+8 D-71} \right]\,.
 \end{align}
Let us summarize the results of this section. For the action \eqref{nmnmCB_v2} we found that conformal gravitational boundary conditions $\delta K= \delta [\beta_{ij}]=0$ are preserved at one loop if the following conditions are met: 1) The boundary condition for the scalar field are of the oblique type of Eq.~$\eqref{scalar_boundary_condition_conformal}$. 2) The non-minimal coupling $\xi$ takes either of the values of Eq.~\eqref{solutions_xi_conformal}. 3) We include in the action an additional minimally coupled complex scalar $\phi'$, of the same mass, with Dirichlet boundary condition $\phi'|_{\pa{\cal M}}=0$.  So while it is possible to preserve the boundary condition, the requirements are certainly bizarre.

\subsection{Maxwell field + conformal boundary}
\label{sec:Maxwell field_conformal boundary}
In this section we examine whether conformal gravitational boundary conditions $\delta[\beta_{ij}]= \delta K=0$ are preserved for the case of a Maxwell field in $D=4$, with Euclidean action:
\beq
\label{EM_action}
I_{\rm EM}= \frac{1}{4}\int_{{\cal M}}d^{D}x \sqrt{g}F_{MN}F^{MN}\,,
\eeq
where $F_{MN}= \pa_{M}A_{N}- \pa_{N}A_{M}$.
In \cite{Jacobson_2014} it was shown that Dirichlet gravitational boundary conditions $\delta\beta_{ij}=0$ are preserved under renormalization for both absolute and relative boundary conditions 
\begin{align}
\begin{split}
\label{Maxwell_relative_absolute_bcs}
A_{n}&|_{\pa{\cal M}}=0\,,~~\nabla_{n}A_{i}+ K_{ij}A^{j} |_{\pa{\cal M}}=0\,,~~ \nabla_{n}\xi|_{\pa{\cal M}}=0\,,\quad\text{(absolute)}\,,\\
A_{i}&|_{\pa{\cal M}}=0\,,~~\nabla_{n}A_{n}+ K A_{n}|_{\pa{\cal M}}=0\,,~~~\xi|_{\pa{\cal M}}=0\,,\quad\text{(relative)}\,,
\end{split}
\end{align}
where the boundary conditions on the gauge parameter $\xi$ are imposed such that the boundary conditions are preserved under the gauge transformation $A_{M}\to A_{M}+ \nabla_{M}\xi$. Because the radiative corrections associated with the boundary conditions~\eqref{Maxwell_relative_absolute_bcs} preserve the gravitational Dirichlet boundary value problem, they do not preserve the conformal gravitational boundary conditions. Thus we consider more general matter boundary conditions of the oblique type
\beq
 \label{Oblique_bc_Maxwell}
A_{n}|_{\pa{\cal M}}=0~~,~~~ [\nabla_{n}A_{m}+ \frac{1}{2}\Gamma^{i}_{mk}D_{i}A^{k}+ \frac{1}{2}D_{i}\left(\Gamma^{i}_{mk}A^{k}\right) + S_{mk}A^{k}]\,|_{\pa{\cal M}}=0\,,
\eeq
with $(\Gamma\indices{^{i}_{mk}}, S_{mk})\in \mathbb{R}$.
Here $A^{k}$ (with raised indices) is defined by raising the components of the pull-back one form $A_{i}$ on the boundary using the inverse boundary metric $\beta^{ij}$. Assuming the Lorentz gauge $\nabla_M A^{M}=0$, the Laplace-type operator \eqref{cal_D_operator} for the Maxwell action \eqref{EM_action} is 
\bes{
\label{Laplace_operator_Maxwell}
{\cal D}=- g^{MN}\nabla_{M}\nabla_{N}+ R_{MN}\,.
}
For the oblique boundary conditions \eqref{Oblique_bc_Maxwell} the operator \eqref{Laplace_operator_Maxwell} is symmetric iff \cite{DMMcAvity_1991}
\beq
\label{symmetry requirements_oblique_Maxwell}
\Gamma\indices{^{i}_{mk}}=-\Gamma\indices{^{i}_{km}},~~S_{mk}=S_{km}\,.
\eeq
Gauge invariance of the oblique BCs \eqref{Oblique_bc_Maxwell}, under $A_{M}\to A_{M}+ \nabla_{M}\xi$ also implies further constraints. If we impose $\nabla_{n}\xi |_{\pa {\cal M}}=0$ and also assume 
\bes{
\label{covariance_of_Gamma}
D_{i}\Gamma\indices{^{i}_{mk}}=0\,,} gauge invariance of Eq.~\eqref{Oblique_bc_Maxwell} implies the following additional conditions on $\Gamma\indices{^{i}_{mk}}$, $S_{mk}$:
\bes{
\label{oblique_under_gauge_variation_v2}
\Gamma\indices{^{(i m)}_{k}}=0~,~ S_{mk}= K_{mk}\,.
}
From Eq.~\eqref{oblique_under_gauge_variation_v2} and Eq.~\eqref{symmetry requirements_oblique_Maxwell} we can deduce that $\Gamma^{imk}$ is a totally antisymmetric tensor. Together with Eq.~\eqref{covariance_of_Gamma} the only possibility for $\Gamma\indices{^{i}_{jk}}$ is 
\bes{
\Gamma\indices{^{i}_{jk}}= a\,\varepsilon\indices{^{i}_{jk}}\,,
}
where $a$ is a constant and $\varepsilon^{ijk}$ is the Levi-Civita (LC) tensor\footnote{The LC tensor is $\varepsilon^{ijk}= (1/\sqrt{\beta})\epsilon^{ijk}$, where $\epsilon^{ijk}$ is the LC symbol and $\beta_{ij}$ the induced metric on $\pa{\cal M}$.}. Note that since this choice of $\Gamma$ does not single out a preferred vector, we cannot form the kinds of boundary scalars that would be needed for gauge loops preserve a microcanonical boundary condition, where the time direction plays a special role.

The oblique-type BCs \eqref{Oblique_bc_Maxwell} with $\Gamma\indices{^{i}_{jk}}=a \varepsilon\indices{^{i}_{jk}}$ and $S_{ij}= K_{ij}$ can be realized by adding the following Chern-Simons (CS)-type boundary term to the Maxwell action \eqref{EM_action}
\bes{
\label{Maxwell_CS_boundary term}
I_{\rm CS}=  -\frac{a}{2}\int_{\pa {\cal M}}d^{3}x\sqrt{\beta} \varepsilon^{ijk}A_{i}D_{j}A_{k}\,.
}
Unfortunately one can not readily use the formulas for the heat kernel coefficients in Eq.~\eqref{coefficientsofHK_heat_kernel} because $[\Gamma^{i},\Gamma^{j}]\neq 0$ in this case. However, the heat kernel coefficients for the CS boundary term \eqref{Maxwell_CS_boundary term} were considered in \cite{Elizalde_1999} for the special case of a 4D-Euclidean ball of radius $r$ with $S^{3}$ boundary. This turns out to be enough to draw a conclusion.
In \cite{Elizalde_1999}, it is shown that the gauge field with CS boundary term on $S^3$   contributes to $a_2$. In the notation of \cite{Elizalde_1999}, 
\begin{equation}
    a_2\sim\frac{1}{(4\pi)^{\frac{3}{2}}}r^{2} C_{1/2},
\end{equation}
where
\begin{equation}
    C_{1/2}=8\pi^{\frac{3}{2}}\frac{1}{a^2-1}.
\end{equation}
When $a=0$, we recover the relative boundary condition. And from \cite{Jacobson_2014}, we know that when $a=0$, integrating out the gauge field with relative boundary condition preserves the Dirichlet boundary condition in the effective action. Now strong ellipticity \cite{Avramidi:1997hy} requires $-1<a<1$ and we observe that  $\left |C_{1/2}(a=0)\right |\leq \left |C_{1/2}(a)\right |$. But in order for the effective action to preserve the conformal boundary condition, we need the absolute value of the coefficient of $\int d^3x\sqrt{\beta} K$ to be smaller than its value at $a=0$. For a 4D-Euclidean ball with $\pa{\cal M}= S^{3}$ only the boundary terms in $a_2$ of Eq.~\eqref{2ndorderofHK_heat_kernel} contribute. Since $K$ is the only boundary scalar that we can form out of $\Gamma\indices{^{i}_{jk}}$ and one power of $K_{ij}$ we conclude that $|C_{1/2}(a=0)|\leq |C_{1/2}(a)|$ implies that the effective action does not preserve the conformal boundary condition in general.

\section{Discussion}

Boundaries in gravitational path integrals provide useful anchors and probes of the gravitational field. Most applications work at leading order in the semiclassical expansion, but various examples point to interesting subtleties when either gravitational or matter field fluctuations are treated. It is already known that not all boundary conditions are compatible with fluctuating gravitational fields, and it seems likely that the same is true for fluctuating matter fields.

It would be interesting to  investigate further the one-loop effective action with non-minimal matter couplings and the microcanonical gravitational boundary conditions, where there is still a puzzle of how to ``untangle" the boundary conditions. We leave this and the extension to more generic matter content to future work.

\section*{Acknowledgements}
PD acknowledges support from the U.S. Department of Energy, Office of Science, Office of High
Energy Physics under award number DE-SC0015655.

~\\
\appendix
\section{Microcanonical boundary conditions}
\label{App:Micro}
\subsection{Minimally coupled complex scalar}
\label{App:A}
In this Appendix we provide detailed derivation of the microcanonical boundary conditions in Euclidean gravity with a minimally coupled massive complex scalar field that satisfies oblique boundary conditions. In $D$-dimensional Euclidean gravity the Einstein-Hilbert action with Gibbons-Hawking-York boundary term is
\beq
\label{Euclidean_EH_GHY}
I_{\text{EH-GHY}}=-\frac{1}{2 \kappa}\int_{{\cal M}} d^{D}x \sqrt{g}\left(R-2 \Lambda\right) - \frac{1}{\kappa}\int_{\pa {\cal M}}\sqrt{\beta}d^{D-1}x K\,,
\eeq
where $K= g^{MN}K_{MN}$ is the trace of the extrinsic curvature tensor of $\pa {\cal M}$ and $\kappa=8\pi G_{N}$. We include the matter action
\beq
\label{matter_action}
\begin{split}
I_{m}&= \int_{{\cal M}}d^{D}x \sqrt{g}\left(g^{MN}\nabla_{M}\phi^{*}\nabla_{N}\phi + m^{2}|\phi|^{2}\right)
+ \int_{\pa{\cal M}}d^{D-1}x\sqrt{\beta} \left[a \phi^{*}\phi + c^{i}\phi^{*}\pa_{i}\phi + c^{* i}\phi \pa_{i}\phi^{*}\right]
\end{split}
\eeq
Here $i,j$ are indices on the boundary $\pa {\cal {M}}$. $a$ and $c^i$ are boundary couplings that will determine the appropriate classical variational problem for $\phi$, and $\beta$ is the codim-1 induced metric on $\partial M$.

The metric and scalar field variation are
\beq
\label{combined_variation_1}
\begin{split}
\delta I_{\text{EH-GHY}}&=-\frac{1}{2\kappa}\int_{{\cal M}} d^{D}x\sqrt{g}\bigg[-R^{MN}+ \frac{1}{2}g^{MN}(R-2\Lambda)\bigg]\delta g_{MN}\\&
- \frac{1}{2\kappa}\int_{\pa {\cal M}}d^{D-1}x \sqrt{\beta}\left((\beta^{ij}K- K^{ij})\delta \beta_{ij}\right)\,,
\end{split}
\eeq
\beq
\label{combined_variation_2}
\begin{split}
&\delta I_{m} =
\frac{1}{2\kappa}\int_{{\cal M}} d^{D}x\sqrt{g}\bigg[ (2\kappa)\left(g^{MN}\left(\frac{1}{2}|\pa \phi |^{2}+ \frac{1}{2}m^{2}|\phi|^{2}\right)+ \nabla^{M}\phi \nabla^{N}\phi^{*}\right)\bigg]\delta g_{MN} \\&
+\int_{{\cal M}}d^{D}x\sqrt{g}\left[\left(-g^{MN}\nabla_{M}\nabla_{N}\phi + m^{2}\phi\right)\delta\phi^{*}+  \left(- g^{MN}\nabla_{M}\nabla_{N}\phi^{*}+ m^{2}\phi^{*}\right)\delta\phi \right]\\&
+ \int_{\pa {\cal M}} d^{D-1}x\sqrt{\beta}\bigg[\left(\frac{1}{2}a |\phi|^{2}+ \frac{1}{2}c^{k}\phi^{*}\pa_{k}\phi + \frac{1}{2}(c^{k})^{*}\phi \pa_{k}\phi^{*}\right)\beta^{ij}\\&\quad\quad\quad\quad\quad\quad\quad~~~+\frac{\pa c^{k}}{\pa \beta_{ij}}\phi^{*}\pa_{k}\phi + \frac{\pa (c^{k})^{*}}{\pa \beta_{ij}}\phi\pa_{k}\phi^{*} \bigg]\delta \beta_{ij}\\&
+\int_{\pa{\cal M}}d^{D-1}x \sqrt{\beta}\left(\nabla_{n}\phi+a\phi -\frac{1}{\sqrt{\beta}}\pa_{i}(\sqrt{\beta}(c^{i})^{*}\phi)+ c^{i}\pa_{i}\phi\right)\delta\phi^{*}\\&
+\int_{\pa{\cal M}}d^{D-1}x \sqrt{\beta}\left(\nabla_{n}\phi^{*}+ a\phi^{*}-\frac{1}{\sqrt{\beta}}\pa_{i}(\sqrt{\beta}c^{i}\phi^{*})+ (c^{i})^{*}\pa_{i}\phi^{*}\right)\delta\phi\,,
\end{split}
\eeq
where boundary total derivatives have been dropped and $\nabla_{n}\phi:= n^{M}\pa_{M}\phi$. In deriving Eq.~\eqref{combined_variation_2} we assumed $a$ is independent of $\beta_{ij}$ while $c^{i}$ depends on the boundary metric $\beta_{ij}$, whose explicit functional dependence will be specified shortly. From the bulk terms we get Einstein's equation and the equation of motion for $\phi$
\beq
\label{eom}
\begin{split}
R^{MN}- \frac{1}{2}g^{MN}R+ \Lambda g^{MN}&=\kappa\left[2\nabla^{(M}\phi \nabla^{N)}\phi^{*}- g^{MN}\left(|\pa \phi |^{2}+ m^{2}|\phi|^{2}\right)\right]\,,\\
g^{MN}\nabla_{M}\nabla_{N}\phi - m^{2}\phi&=0\,.
\end{split}
\eeq

Let us now discuss boundary conditions. From \eqref{combined_variation_1},\eqref{combined_variation_2} we see that the boundary terms vanish when
\beq
\delta \beta_{ij}\big |_{\pa {\cal M}}=0~~,~~ \left[\nabla_{n}\phi+a\phi -\frac{1}{\sqrt{\beta}}\pa_{i}(\sqrt{\beta}c^{i*}\phi)+ c^{i}\pa_{i}\phi\right]\big|_{\pa{\cal M}}=0\,.
\eeq
To match with the oblique boundary conditions given in Eq.~\eqref{GammaSbc}
\begin{align}
    [\nabla_n+\frac{1}{2}(\Gamma^i D_i+D_i\Gamma^i)+S]\phi=0\,,
    \label{GammaSbc_app}
\end{align}
we simply identify
\beq
a=S~~,~~c^{i}= \frac{\Gamma^{i}}{2}~~,~~\text{Re}(\Gamma^{i})=\text{Re}(c^{i})=0\,.
\eeq

Let us now turn to the microcanonical ensemble. The usual procedure for pure gravity involves adding appropriate boundary terms\footnote{This is what we meant by a ``Legendre transformation" in Section~\ref{subsec:Formalism and the example of scalar field}.} to the action $I_{\text{EH-GHY}}$ such that the lapse and shift of the boundary metric are unconstrained in the variational problem. The covariant form of the mircrocanonical action in pure Gravity was given in~\cite{PhysRevD.47.1420}\,,
\beq
\label{EH-micro}
I_{\text{EH-micro}}= - \frac{1}{2\kappa}\int_{{\cal M}}d^{D}x \sqrt{g}(R- 2\Lambda) - \frac{1}{\kappa}\int_{t_{1},t_{2}}d^{D-1}x\sqrt{\beta}{\cal K} - \frac{1}{\kappa}\int_{{\cal B}\times I}d^{D-1}x\sqrt{\beta}t_{i}K^{ij}\pa_{j}t
\eeq
where $t_{1},t_{2}$ denote the constant ``spacelike" parts of the boundary with extrinsic trace ${\cal K}$, and  ${\cal B}\times I$ denotes the ``timelike" boundary and its extrinsic tensor $K_{ij}$. $\un t= \un\pa_{t}$ is the time vector for the time foliation $t$ of the boundary and we will denote $\un u$ to be the unit normal vector of the time foliation. Note that because of the orthogonality condition (which we assume) we have
\beq
u_{M}n^{M} |_{{\cal B}\times I}=0\,,
\eeq
which implies $\un u |$ is tangent to the boundary ${\cal B}\times I$. 

The lapse, shift and codim-2 metric $N, V^{a}, s_{ab}$ are defined through the ADM decomposition of the boundary metric $\beta_{ij}$ on ${\cal B}\times I$, namely
\beq
\label{boundary_beta_ADM_decomposition}
\beta_{ij}dx^{i}dx^{j}= Ndt^{2}+ s_{ab}\left(dx^{a}+ V^{a}dt\right)\left(dx^{b}+ V^{b}dt\right)\,.
\eeq
It is also useful to define the BY surface currents $\varepsilon_{BY}, J_{a}, J^{ab}$ 
\beq
\label{epsilon_J_Jab_definitions}
\varepsilon_{BY}= -u_{i}u_{j}\tau^{ij}~~,~~ J_{a}= \sigma_{ai}u_{j}\tau^{ij}~~,~~J^{ab}=\sigma^{a}_{i}\sigma^{b}_{j}\tau^{ij}\,,
\eeq
where $\sigma_{ij}=\beta_{ij}-u_{i}u_{j}$,
\beq
\label{BY_stress}
\tau^{ij}= \frac{2}{\sqrt{\beta}}\pi^{ij}~~~,~~~\pi^{ij}= \frac{1}{2\kappa}\sqrt{\beta}\left(K^{ij}- K \beta^{ij}\right)\,,
\eeq
and $K^{ij}$ is the extrinsic curvature tensor of the timelike boundary ${\cal B}\times I$.
By employing a Gauss-Codazzi  type of decomposition of $K^{ij}$ one can show that BY surface energy density $\varepsilon_{BY}$ in \eqref{epsilon_J_Jab_definitions} can be expressed as
\beq
\label{varepsilon_expression_cod_2_k}
\varepsilon_{BY}=\frac{1}{\kappa }k\,,
\eeq
with $k$ is the trace of the extrinsic curvature of the codimension-2 boundary ${\cal B}$. We emphasize that the above form of the BY surface density holds only in pure gravity without matter fields.

In terms of the ADM variables the microcanonical boundary terms in Eq.~\eqref{EH-micro} can be expressed as follows
\beq
\label{Micro_boundary_term_ADM}
\begin{split}
- \frac{1}{\kappa}\int_{{\cal B}\times I}d^{D-1}x\sqrt{\beta}t_{i}K^{ij}\pa_{j}t = -\frac{1}{\kappa}\int_{{\cal B}\times I}d^{D-1}x \sqrt{\beta}\left(K- k\right)- \int_{{\cal B}\times I}d^{D-1}x \sqrt{s}V^{a}J_{a}\,.
\end{split}
\eeq
In Section~\ref{subsec:Formalism and the example of scalar field} we gauged fixed to $V^{a}|_{{\cal B}\times I}=0$, and only the $[K-k]$ structure remains in the microcanonical boundary.

The variation of $I_{\text{EH-micro}}$ is
\beq
\label{EH-micro_variation}
\begin{split}
\delta I_{\text{EH-micro}}=&-\frac{1}{2\kappa}\int d^{D}x\sqrt{g}\bigg[-R^{MN}+ \frac{1}{2}g^{MN}(R-2\Lambda)\bigg]\delta g_{MN} \\&
+\int_{t_{1},t_{2}}d^{D-1}x \sqrt{\beta}P^{ij}\delta\beta_{ij}\\&
+ \int_{{\cal B}\times I}d^{D-1}x \left[N\delta(\sqrt{s}\varepsilon_{BY}) - V^{a}\delta(\sqrt{s}J_{a})+ \frac{1}{2}N\sqrt{s}J^{ab}\delta s_{ab} \right]\,,
\end{split}
\eeq
where $P^{ij}= \frac{1}{2\kappa}\left({\cal K}^{ij}- {\cal K} \beta^{ij}\right)$ with ${\cal K}^{ij}$ the extrinsic curvature tensor of the time slices $t_{1},t_{2}$.

Let us now consider the variation of the matter action. We will focus on the following $\beta_{ij}$ dependence of $c^{i}$
\beq
\label{c_i_ansatz}
\un c=i \un u=iN^{-1}(\un t - \un V)\,,
\eeq
where $\un u$ is the unit normal to the time foliation and $\un V$ the shift vector.

Similar to pure gravity, the microcanonical ensemble is defined by adding appropriate boundary terms that such that the lapse and shift boundary variations are unconstrained. From $\delta I_{m}$ we see that the relevant boundary variations are contained in the third line of \eqref{combined_variation_2} for the timelike boundary ${\cal B}\times I$
\beq
\label{blue_term_variation_case_1}
\begin{split}
&\int_{{\cal B}\times I} d^{D-1}x\sqrt{\beta}\bigg[\left(\frac{1}{2}a |\phi|^{2}+ \frac{1}{2}c^{k}\phi^{*}\pa_{k}\phi + \frac{1}{2}(c^{k})^{*}\phi \pa_{k}\phi^{*}\right)\beta^{ij} \\&
\quad\quad\quad\quad\quad\quad\quad+\frac{\pa c^{k}}{\pa \beta_{ij}}\phi^{*}\pa_{k}\phi + \frac{\pa (c^{k})^{*}}{\pa \beta_{ij}}\phi\pa_{k}\phi^{*} \bigg]\delta \beta_{ij} =\\&\int_{{\cal B}\times I}d^{D-1}x \sqrt{s}\bigg[ a|\phi|^{2}\delta N  + i (\phi\pa_{a}\phi^{*}- \phi^{*}\pa_{a}\phi)\delta V^{a} \\&
\quad\quad\quad\quad\quad\quad\quad
+ \frac{1}{2}N\left(a|\phi|^2 +c^{k}\phi^{*}\pa_{k}\phi + (c^{k})^{*}\phi\pa_{k}\phi^{*}\right)s^{ab}\delta s_{ab}
\bigg]\,,
\end{split}
\eeq
where we used \eqref{c_i_ansatz} and the following useful relations
\begin{align}
\label{beta_variation_ADM_1}
\delta \beta_{ij}&= 2N^{-1}u_{i}u_{j}\delta N  + 2 N^{-1} \sigma_{a( i}u_{j)}\delta V^{a}+ \sigma^{a}_{(i}\sigma^{b}_{j)}\delta s_{ab}\,,\\\label{beta_variation_ADM_2}
\beta^{ij}\delta\beta_{ij}&=2N^{-1}\delta N + s^{ab}\delta s_{ab}\,,\\\label{beta_variation_ADM_3}
\frac{\pa c^{k}}{\pa \beta_{ij}}\delta \beta_{ij}&=- N^{-1} c^{k}\delta N - iN^{-1}\delta^{k}_{a} \delta V^{a}\,.
\end{align}

From \eqref{blue_term_variation_case_1} and \eqref{c_i_ansatz} we find that the appropriate boundary term needed to shift to the microcanonical ensemble is
\begin{align}
\label{S_boundary_case_1}
S_{\text{boundary,micro}}=-\int_{{\cal B}\times I}d^{D-1}x \sqrt{\beta}a |\phi|^{2} -\int_{{\cal B}\times I}d^{D-1}x \sqrt{\beta} N^{-1}\left[i\left(\phi\pa_{a}\phi^{*}- \phi^{*}\pa_{a}\phi\right)V^{a}\right]\,.
\end{align}
Adding the boundary term \eqref{S_boundary_case_1} into $I_{m}$ given in \eqref{matter_action} and using \eqref{EH-micro} we find the (total) microcanonical action
\beq
\label{micro_action_ansatz_1}
\begin{split}
&I_{\text{micro}}:= I_{\text{EH-micro}} + I_{m}+ S_{\text{boundary,micro}}\\&
=- \frac{1}{2\kappa}\int_{{\cal M}}d^{D}x \sqrt{g}(R- 2\Lambda) - \frac{1}{\kappa}\int_{t_{1},t_{2}}d^{D-1}x\sqrt{\beta}{\cal K} - \frac{1}{\kappa}\int_{{\cal B}\times I}d^{D-1}x\sqrt{\beta}t_{i}K^{ij}\pa_{j}t \\&
+\int_{{\cal M}}d^{D}x \sqrt{g}\left(g^{MN}\nabla_{M}\phi^{*}\nabla_{N}\phi + m^{2}|\phi|^{2}\right)+ \int_{t_{1},t_{2}}d^{D-1}x\sqrt{\beta} \left[a \phi^{*}\phi + i u^{i}\phi^{*}\pa_{i}\phi - i\phi u^{i} \pa_{i}\phi^{*}\right]\\&
+\int_{{\cal B}\times I}d^{D-1}x \sqrt{\beta}N^{-1}i t^{i}\left[\phi^{*}\pa_{i}\phi - \phi\pa_{i}\phi^{*}  \right]\,.
\end{split}
\eeq
Setting $V^{a}|_{{\cal B}\times I}=0$ and making use of \eqref{Micro_boundary_term_ADM} we can see the relevant timelike boundary terms of $I_{\text{micro}}$ above reduce to the Eq.~\eqref{Imicro2} given in Section~\ref{subsec:Formalism and the example of scalar field}. 

Finally, if we calculate the variation of $I_{\text{micro}}$ we obtain
\beq
\label{Variation_of_total_microcanonical_action}
\begin{split}
&\delta I_{\text{micro}} = -\frac{1}{2\kappa}\int d^{D}x\sqrt{g}\bigg[-R^{MN}+ \frac{1}{2}g^{MN}(R-2\Lambda)\bigg]\delta g_{MN} \\&
+\int_{t_{1},t_{2}}d^{D-1}x \sqrt{\beta}P^{ij}\delta\beta_{ij}
+\int_{{\cal M}}d^{D}x \left[g^{MN}\left(\frac{1}{2}|\pa \phi |^{2}+ \frac{1}{2}m^{2}|\phi|^{2}\right) - \nabla^{M}\phi \nabla^{N}\phi^{*}\right]\delta g_{MN}\\&
+\int_{{\cal M}}d^{D}x\sqrt{g}\left(\left(-g^{MN}\nabla_{M}\nabla_{N}\phi + m^{2}\phi\right)\delta\phi^{*}+  \left(- g^{MN}\nabla_{M}\nabla_{N}\phi^{*}+ m^{2}\phi^{*}\right)\delta\phi \right)+\\&
\int_{t_{1},t_{2}} d^{D-1}x\sqrt{\beta}\bigg[\left(\frac{1}{2}a |\phi|^{2}+ \frac{1}{2}c^{k}\phi^{*}\pa_{k}\phi + \frac{1}{2}(c^{k})^{*}\phi \pa_{k}\phi^{*}\right)\beta^{ij}+\frac{\pa c^{k}}{\pa \beta_{ij}}\phi^{*}\pa_{k}\phi + \frac{\pa (c^{k})^{*}}{\pa \beta_{ij}}\phi\pa_{k}\phi^{*} \bigg]\delta \beta_{ij}\\&
+\int_{\pa{\cal M}}d^{D-1}x \sqrt{\beta}\left(\nabla_{n}\phi+a\phi -\frac{1}{\sqrt{\beta}}\pa_{i}(\sqrt{\beta}(c^{i})^{*}\phi)+ c^{i}\pa_{i}\phi\right)\delta\phi^{*}\\&
+\int_{\pa{\cal M}}d^{D-1}x \sqrt{\beta}\left(\nabla_{n}\phi^{*}+ a\phi^{*}-\frac{1}{\sqrt{\beta}}\pa_{i}(\sqrt{\beta}c^{i}\phi^{*})+ (c^{i})^{*}\pa_{i}\phi^{*}\right)\delta\phi\\&
+ \int_{{\cal B}\times I}d^{D-1}x \bigg[N\delta[\sqrt{s}(\varepsilon_{BY} - a |\phi|^{2})] - V^{a}\delta[\sqrt{s}\left(J_{a}+ i(\phi\pa_{a}\phi^{*}-\phi^{*}\pa_{a}\phi)\right)] \\&
+ \frac{1}{2}N\sqrt{s}\left(J^{ab}+ \left(a|\phi|^2 +c^{k}\phi^{*}\pa_{k}\phi + (c^{k})^{*}\phi\pa_{k}\phi^{*}\right)s^{ab}\right)\delta s_{ab} \bigg]\,,
\end{split}
\eeq
where $c^{k}= i u^{k}$ (see Eq.~\eqref{c_i_ansatz}) and $\frac{\pa c^{k}}{\pa \beta_{ij}}\delta \beta_{ij}=- N^{-1}\left(c^{k}\delta N + i\delta^{k}_{a}\delta V^{a}\right)$.

From Eq.~\eqref{Variation_of_total_microcanonical_action} we can read off how the new BY surface currents due to the presence of matter
\beq
\begin{split}
\varepsilon_{\text{new}}:= \varepsilon_{BY} - a |\phi|^{2}~~&,~~J^{\text{new}}_{a}:=J_{a}+ i(\phi\pa_{a}\phi^{*}-\phi^{*}\pa_{a}\phi)\,, \\J_{\text{new}}^{ab}:=J^{ab}+&\left(a|\phi|^2 + c^{k}\phi^{*}\pa_{k}\phi + (c^{k})^{*}\phi\pa_{k}\phi^{*}\right)s^{ab}
\end{split}
\eeq
where $\varepsilon_{BY}, J_{a}, J^{ab}$ were defined in \eqref{epsilon_J_Jab_definitions}.

\subsection{Non-minimally coupled complex scalar}
\label{App:B}
In this Appendix we consider the case of a non-minimally coupled massive complex scalar with action 
\begin{align}
\label{non_minimal_matter_action}
    I_{\rm n.m.}=&\int_{\cal M} d^Dx\sqrt{g}\left(\vert\nabla\phi\vert^2+\phi^*(m^2-\xi R)\phi \right)\nonumber\\
    &+\int_{\partial {\cal M}}d^{D-1}x\sqrt{\beta}\left(-2K \xi|\phi|^2+ a|\phi|^2 +c^{i}\phi^*\partial_i\phi+ c^{i*}\phi \pa_{i}\phi^{*}\right)\,,
\end{align}
where $\xi$ is the non-minimal (real) coupling constant.
We consider the same Einstein-Hilbert action $I_{\text{EH-GHY}}$ as in Eq.~\eqref{Euclidean_EH_GHY} with metric variation given in Eq.~\eqref{combined_variation_1}. Taking the variation of \eqref{non_minimal_matter_action} we obtain\footnote{We point out that when $\pa{\cal M}$ refers to the spacelike part of the boundary $K$ should be replaced by ${\cal K}$ according to our conventions. See below Eq.~\eqref{EH-micro_variation}.}
\bes{
\label{non_minimal_total_variation}
&\delta I_{\rm n.m.}=\int_{{\cal M}}d^{D}x \sqrt{g}\bigg[- \nabla^{M}\phi\nabla^{N}\phi^{*} + \frac{1}{2}g^{MN}\left(\vert\pa_{M}\phi\vert^2+\phi^*(m^2-\xi R)\phi\right)\\& + \xi |\phi|^2 R^{MN}
+ \xi \left(g^{MN}\nabla_{K}\nabla^{K}- \nabla^{M}\nabla^{N}\right)|\phi|^{2}\bigg]\delta g_{MN}\\&
+\int_{\cal M}d^{D}x\sqrt{g}\left[-g^{MN}\nabla_{M}\nabla_{N}\phi^{*}+ (m^{2}-\xi R)\phi^{*}\right]\delta\phi \\&
+ \int_{{\cal M}}d^{D}x\sqrt{g}\left[-g^{MN}\nabla_{M}\nabla_{N}\phi+ (m^{2}-\xi R)\phi\right]\delta\phi^{*}\\&
+\int_{\pa {\cal M}}d^{D-1}x\sqrt{\beta}\bigg[-\xi\beta^{ij}\nabla_{n}|\phi|^2 +\frac{\pa c^{k}}{\pa \beta_{ij}}\phi^{*}\pa_{k}\phi +\frac{\pa (c^{k})^{*}}{\pa \beta_{ij}}\phi\pa_{k}\phi^{*} \\&\quad\quad\quad\quad + \frac{1}{2}\beta^{ij}\left(a|\phi|^2 +c^{k}\phi^*\partial_k\phi+ (c^{k})^{*}\phi \pa_{k}\phi^{*}\right) + \xi |\phi|^2(K^{ij}- K \beta^{ij})
\bigg]\delta\beta_{ij}\\&
+\int_{\pa{\cal M}}d^{D-1}x\sqrt{\beta}\left[\nabla_{n}\phi^{*} + a\phi^{*}- 2K\xi\phi^{*}- \frac{1}{\sqrt{\beta}}\pa_{i}(\sqrt{\beta}c^{i}\phi^{*}) + (c^{i})^{*}\pa_{i}\phi^{*}\right]\delta\phi \\&
+\int_{\pa{\cal M}}d^{D-1}x\sqrt{\beta}\left[\nabla_{n}\phi + a\phi - 2K \xi \phi- \frac{1}{\sqrt{\beta}}\pa_{i}(\sqrt{\beta}(c^{i})^{*}\phi) + c^{i}\pa_{i}\phi\right]\delta\phi^{*}\,.
}
In the above, we assumed that $c^{i}$ depends on the boundary metric $\beta_{ij}$ while $a$ is boundary metric independent.

From the bulk terms we get Einstein's equation and the equation of motion for $\phi$
\beq
\begin{split}
R^{MN}- \frac{1}{2}g^{MN}R+ \Lambda g^{MN}&=\kappa\left[2\nabla^{(M}\phi \nabla^{N)}\phi^{*} - 2\xi\left(g^{MN}\nabla_{K}\nabla^{K}|\phi|^{2}- \nabla^{M}\nabla^{N}|\phi|^{2}\right)\right] \\&~~~~~~-2\kappa\xi R^{MN}|\phi|^{2} - \kappa g^{MN}\left(|\pa \phi |^{2}+ \phi^{*}(m^{2}-\xi R)\phi\right)\,,\\
g^{MN}\nabla_{M}\nabla_{N}\phi - \left(m^{2}-\xi R\right)\phi&=0\,,
\end{split}
\eeq
where $\kappa=8\pi G_{N}$. Let us now discuss boundary conditions. From Eqs.~\eqref{combined_variation_1},\eqref{non_minimal_total_variation} we see that the boundary terms vanish when
\bes{
\delta \beta_{ij}|_{\pa{\cal M}}=0~~,~~\nabla_{n}\phi + a\phi - 2K \xi \phi- \frac{1}{\sqrt{\beta}}\pa_{i}(\sqrt{\beta}c^{i*}\phi) + c^{i}\pa_{i}\phi |_{\pa {\cal M}}=0\,.
}
Let us now turn to the microcanonical ensemble for the non-minimal case. The microcanonical action for the Einstein-Hilbert part is the same as the minimally coupled case and was given in Eq.~\eqref{EH-micro} with variation obtained in Eq.~\eqref{EH-micro_variation}. 

For the matter part in Eq.~\eqref{non_minimal_matter_action} new boundary terms are needed to have unconstrained lapse and shift variations. For this a specific $\beta_{ij}$ dependence of $c^{i}$ is required. We will assume the same ansatz as before, namely Eq.~\eqref{c_i_ansatz}. From $\delta I_{n.m.}$ we see that the relevant boundary variations are contained in the fifth line of \eqref{non_minimal_total_variation} for the timelike boundary ${\cal B}\times I$
\bes{
&\int_{{\cal B}\times I}d^{D-1}x\sqrt{\beta}\bigg[-\xi\beta^{ij}\nabla_{n}|\phi|^2 +\frac{\pa c^{k}}{\pa \beta_{ij}}\phi^{*}\pa_{k}\phi +\frac{\pa (c^{k})^{*}}{\pa \beta_{ij}}\phi\pa_{k}\phi^{*} \\&\quad\quad\quad\quad + \frac{1}{2}\beta^{ij}\left(a|\phi|^2 +c^{k}\phi^*\partial_k\phi+ (c^{k})^{*}\phi \pa_{k}\phi^{*}\right) + \xi |\phi|^2(K^{ij}- K \beta^{ij})
\bigg]\delta\beta_{ij}\\&
=\int_{{\cal B}\times I}d^{D-1}x\sqrt{s}\bigg[a|\phi|^{2}- 2\kappa\xi |\phi|^{2} \varepsilon_{BY} -2\xi \nabla_{n}(|\phi|^2) \bigg]\delta N\\&~~
+\int_{{\cal B}\times I}d^{D-1}x\sqrt{s}\left[2\kappa \xi |\phi|^{2} J_{a}+ i (\phi\pa_{a}\phi^{*}-  \phi^{*}\pa_{a}\phi)\right]\delta V^{a}\\&~~
+\int_{{\cal B}\times I}d^{D-1}x\sqrt{s}\left[N\kappa \xi |\phi|^{2} J^{ab}+ \frac{1}{2}N\left(a|\phi|^2 -2\xi \nabla_{n}(|\phi|^{2}) +c^{i}\phi^*\partial_i\phi+ (c^{i})^{*}\phi \pa_{i}\phi^{*}\right)s^{ab} \right]\delta s_{ab}\,,
}
where we used the ansatz \eqref{c_i_ansatz} and the formulas in Eqs~\eqref{beta_variation_ADM_2}-\eqref{beta_variation_ADM_3} and the definition of $\varepsilon_{BY}$ given in Eq. \eqref{varepsilon_expression_cod_2_k}. 

To remove the lapse and shift variation we need to add the following boundary term
\begin{align}
\label{S_boundary_case_non_minimal}
S_{\text{boundary-n.m.}}&=-\int_{{\cal B}\times I}d^{D-1}x \sqrt{\beta}\bigg[a|\phi|^{2}- 2\kappa\xi |\phi|^{2} \varepsilon_{BY} -2\xi \nabla_{n}(|\phi|^2) \bigg]\\&
 -\int_{{\cal B}\times I}d^{D-1}x \sqrt{\beta} N^{-1}\left[2\kappa \xi |\phi|^{2} J_{a}V^{a}+i\left(\phi\pa_{a}\phi^{*}- \phi^{*}\pa_{a}\phi\right)V^{a}\right]\,.
\end{align}
The  microcanonical action for the non-minimal case is then
\beq
\label{micro_action_non_minimal}
\begin{split}
&I_{\text{micro-n.m.}}:= I_{\text{EH-micro}} + I_{\rm n.m.}+ S_{\text{boundary-n.m.}}\\&
=- \frac{1}{2\kappa}\int_{{\cal M}}d^{D}x \sqrt{g}(R- 2\Lambda) - \frac{1}{\kappa}\int_{t_{1},t_{2}}d^{D-1}y\sqrt{\beta}{\cal K} - \frac{1}{\kappa}\int_{{\cal B}\times I}d^{D-1}y\sqrt{\beta}t_{i}K^{ij}\pa_{j}t \\&
+\int_{\cal M} d^Dx\sqrt{g}\left(\vert\pa_{M}\phi\vert^2+\phi^*(m^2-\xi R)\phi \right)\\&
 +\int_{t_{1},t_{2}}d^{D-1}x\sqrt{\beta}\left(-2{\cal K} \xi|\phi|^2+ a|\phi|^2 +c^{i}\phi^*\partial_i\phi+ c^{i*}\phi \pa_{i}\phi^{*}\right)\\&
 +\int_{{\cal B}\times I}d^{D-1}x\sqrt{\beta}\left[2(k- K)\xi |\phi|^{2}+ 2\xi \nabla_{n}(|\phi|^2)\right]\\&
 +\int_{{\cal B}\times I}d^{D-1}x\sqrt{s}\left[-2\kappa\xi |\phi|^{2}J_{a}V^{a}+ it^{i}(\phi^{*}\pa_{i}\phi -\phi\pa_{i}\phi^{*})\right]\,,
\end{split}
\eeq
with variation
\bes{
\label{micro_variation_non_minimal}
&\delta I_{\text{micro-non-minimal}}= -\frac{1}{2\kappa}\int d^{D}x\sqrt{g}\bigg[-R^{MN}+ \frac{1}{2}g^{MN}(R-2\Lambda)\bigg]\delta g_{MN} +\int_{t_{1},t_{2}}d^{D-1}y P^{ij}\delta\beta_{ij}\\&
+\int_{{\cal M}}d^{D}x \sqrt{g}\bigg[- \nabla^{M}\phi\nabla^{N}\phi^{*} + \frac{1}{2}g^{MN}\left(\vert\pa_{M}\phi\vert^2+\phi^*(m^2-\xi R)\phi\right)+ \xi |\phi|^2 R^{MN} \\&
+ \xi \left(g^{MN}\nabla_{K}\nabla^{K}- \nabla^{M}\nabla^{N}\right)|\phi|^{2}\bigg]\delta g_{MN}\\&
+\int_{t_{1},t_{2}}d^{D-1}x\sqrt{\beta}\bigg[-\xi\beta^{ij}\nabla_{n}|\phi|^2 +\frac{\pa c^{k}}{\pa \beta_{ij}}\phi^{*}\pa_{k}\phi +\frac{\pa (c^{k})^{*}}{\pa \beta_{ij}}\phi\pa_{k}\phi^{*}+ \xi |\phi|^2({\cal K}^{ij}- {\cal K} \beta^{ij}) \\&
+ \frac{1}{2}\beta^{ij}\left(a|\phi|^2 +c^{k}\phi^*\partial_k\phi+ (c^{k})^{*}\phi \pa_{k}\phi^{*}\right)
\bigg]\delta\beta_{ij} + \int_{{\cal M}}d^{D}x\sqrt{g}\left[-g^{MN}\nabla_{M}\nabla_{N}\phi+ (m^{2}-\xi R)\phi\right]\delta\phi^{*}\\&
+\int_{\cal M}d^{D}x\sqrt{g}\left[-g^{MN}\nabla_{M}\nabla_{N}\phi^{*}+ (m^{2}-\xi R)\phi^{*}\right]\delta\phi \\&
+\int_{\pa{\cal M}}d^{D-1}x\sqrt{\beta}\left[\nabla_{n}\phi^{*} + a\phi^{*}- 2K\xi\phi^{*}- \frac{1}{\sqrt{\beta}}\pa_{i}(\sqrt{\beta}c^{i}\phi^{*}) + (c^{i})^{*}\pa_{i}\phi^{*}\right]\delta\phi \\&
+\int_{\pa{\cal M}}d^{D-1}x\sqrt{\beta}\left[\nabla_{n}\phi + a\phi - 2K \xi \phi- \frac{1}{\sqrt{\beta}}\pa_{i}(\sqrt{\beta}(c^{i})^{*}\phi) + c^{i}\pa_{i}\phi\right]\delta\phi^{*}\\&
+\int_{{\cal B}\times I}d^{D-1}x \frac{1}{2} N\sqrt{s}\left[J^{ab}+ 2\kappa \xi |\phi|^{2} J^{ab}+ \left(a|\phi|^2 -2\xi \nabla_{n}(|\phi|^{2}) +c^{i}\phi^*\partial_i\phi+ (c^{i})^{*}\phi \pa_{i}\phi^{*}\right)s^{ab} \right]\delta s_{ab}\\&
+\int_{{\cal B}\times I}d^{D-1}x N \delta\left(\sqrt{s}\bigg[\varepsilon_{BY} - a|\phi|^{2}+ 2\kappa\xi |\phi|^{2} \varepsilon_{BY} + 2\xi \nabla_{n}(|\phi|^2) \bigg]\right)\\&
 -\int_{{\cal B}\times I}d^{D-1}x V^{a}\delta \left( \sqrt{s}\left[J_{a}+ 2\kappa \xi |\phi|^{2} J_{a}+i\left(\phi\pa_{a}\phi^{*}- \phi^{*}\pa_{a}\phi\right)\right]\right)\,,
}
where $P^{ij}$ was defined below $\eqref{EH-micro_variation}$, $c^{k}= i u^{k}$ (see Eq.~\eqref{c_i_ansatz}) and $\frac{\pa c^{k}}{\pa \beta_{ij}}\delta \beta_{ij}=- N^{-1}\left(c^{k}\delta N + i\delta^{k}_{a}\delta V^{a}\right)$.
From the last three lines of Eq.~\eqref{micro_variation_non_minimal} we see that the BY currents are shifted as follows
\bes{
\varepsilon_{\text{new}}:&= \varepsilon_{BY}(1+2\kappa\xi |\phi|^2)+ 2\xi \nabla_{n}(|\phi|^2) - a|\phi|^{2}\\
J^{\text{new}}_{a}:&=J_{a}(1+ 2\kappa \xi |\phi|^2)+ i(\phi\pa_{a}\phi^{*}-\phi^{*}\pa_{a}\phi)\\
J_{\text{new}}^{ab}:&=J^{ab}(1+ 2\kappa \xi |\phi|^2)+\left(a|\phi|^2 -2\xi\nabla_{n}|\phi|^2 + c^{k}\phi^{*}\pa_{k}\phi + (c^{k})^{*}\phi\pa_{k}\phi^{*}\right)s^{ab}
}
where $\varepsilon_{BY}, J_{a}, J^{ab}$ where defined in \eqref{epsilon_J_Jab_definitions}.

\section{Non-minimally coupled complex scalar + conformal boundary}
\label{App:C}
In this Appendix we consider the case of a non-minimally coupled massive complex scalar with conformal gravitational boundary conditions. 

The gravitational action with conformal bc's is
\beq
\label{Euclidean_EH_GHY_conformal}
I_{\rm{EH-CGHY}}=-\frac{1}{2 \kappa}\int d^{D}x \sqrt{g}\left(R-2 \Lambda\right) - \frac{1}{(D-1)}\frac{1}{\kappa}\int_{\pa {\cal M}}\sqrt{\beta}d^{D-1}y K\,.
\eeq
Taking the variation of Eq.~\eqref{Euclidean_EH_GHY_conformal} we find
\bes{
\label{EH_CGHY_variation}
\delta I_{\rm{EH-CGHY}}&=\frac{1}{2\kappa}\int d^{D}x\sqrt{g}\bigg[-R^{MN}+ \frac{1}{2}g^{MN}(R-2\Lambda)\bigg]\delta g_{MN}\\&
-\frac{1}{2\kappa}\int_{\pa{\cal M}}d^{D-1}y \sqrt{\beta}\left(\frac{1}{D-1}K \beta^{ij}- K^{ij}\right)\delta\beta_{ij} +  \left(\frac{D-2}{D-1}\right)\frac{1}{\kappa}\int_{\pa{\cal M}}d^{D-1}y \sqrt{\beta}\delta K\,.
}
We see that in order for the first term in the second line above to vanish, only the conformal structure of $[\beta_{ij}$] needs to be fixed, since the boundary tensor $T^{ij}\equiv (1/(D-1))K \beta^{ij}- K^{ij}$ is traceless, namely $\beta_{ij} T^{ij}=0$. Therefore, the boundary terms in Eq.~\eqref{EH_CGHY_variation} vanish if we impose $\delta[\beta_{ij}]=\delta K=0$.

For the matter part, the action is 
\bes{
\label{conformal_matter_action}
    I_{\rm{m.conf}}=&\int_{\cal M} d^Dx\sqrt{g}\left(\vert\nabla\phi\vert^2+\phi^*(m^2-\xi R)\phi \right) +\int_{\partial {\cal M}}d^{D-1}y\sqrt{\beta}\left(-2\frac{1}{D-1}K \xi|\phi|^2 - b|\phi|^2 \right)\\&- \int_{\pa{\cal M}}d^{D-1}y\sqrt{\beta}(a K |\phi|^{2})\,,
}
with $a,b$ constants. The variation of Eq.~\eqref{conformal_matter_action} is
\bes{
\label{conformal_matter_variation}
&\delta I_{\rm{m.conf}}= \int_{{\cal M}}d^{D}x \sqrt{g}\bigg[- \nabla^{M}\phi\nabla^{N}\phi^{*} + \frac{1}{2}g^{MN}\left(\vert\nabla\phi\vert^2+\phi^*(m^2-\xi R)\phi\right)+ \xi |\phi|^2 R^{MN} \\&
+ \xi \left(g^{MN}\nabla_{K}\nabla^{K}- \nabla^{M}\nabla^{N}\right)|\phi|^{2}\bigg]\delta g_{MN}\\&
+\int_{\pa {\cal M}}d^{D-1}x\sqrt{\beta}\bigg[-\xi\beta^{ij}\nabla_{n}|\phi|^2 - \frac{1}{2}\beta^{ij}b|\phi|^2 
+ \xi |\phi|^2(K^{ij}- \frac{1}{D-1}K \beta^{ij}) - \frac{1}{2}a |\phi|^{2}K\beta^{ij}
\bigg]\delta\beta_{ij} \\&
+\int_{\pa{\cal M}}d^{D-1}x |\phi|^{2}\left[2\xi\left(\frac{D-2}{D-1}\right)- a\right] \delta K  \\&
+\int_{\cal M}d^{D}x\sqrt{g}\left[-g^{MN}\nabla_{M}\nabla_{N}\phi^{*}+ (m^{2}-\xi R)\phi^{*}\right]\delta\phi \\&
+ \int_{{\cal M}}d^{D}x\sqrt{g}\left[-g^{MN}\nabla_{M}\nabla_{N}\phi+ (m^{2}-\xi R)\phi\right]\delta\phi^{*}\\&
+\int_{\pa{\cal M}}d^{D-1}x\sqrt{\beta}\left[\nabla_{n}\phi^{*} - b\phi^{*}- \frac{2}{D-1}K\xi\phi^{*}- a K\phi^{*}\right]\delta\phi \\&
+\int_{\pa{\cal M}}d^{D-1}x\sqrt{\beta}\left[\nabla_{n}\phi - b\phi - \frac{2}{D-1}K \xi \phi - a K\phi\right]\delta\phi^{*}
}

The total action $I_{{\rm conf}}=I_{\rm{EH-CGHY}}+ I_{\rm{m.conf}}$ (given in Eq.~\eqref{nmnmCB_v2} in the main text for $\Lambda=0$) is
\begin{align}
    \label{nmnmCB_v2_App}
     I_{\rm conf}=&\int_M d^Dx\sqrt{g}\left(\vert\nabla\phi\vert^2+\phi^*(m^2-\xi R)\phi-\frac{1}{2\kappa}(R-2\Lambda)\right)\nonumber\\
    -&\frac{1}{\kappa}\int_{\partial M}d^{D-1}x\sqrt{\beta} \left(\frac{1}{D-1}K(1+2\kappa\xi\vert\phi\vert^2)+\kappa(aK+b)|\phi|^2\right)\,.
\end{align}
Adding \eqref{EH_CGHY_variation} and \eqref{conformal_matter_variation} we obtain for the total variation of \eqref{nmnmCB_v2_App}
\bes{
\label{conformal_total_variation_paper_v1}
&\delta I_{\rm conf} =\frac{1}{2\kappa}\int d^{D}x\sqrt{g}\bigg[-R^{MN}+ \frac{1}{2}g^{MN}(R-2\Lambda)\bigg]\delta g_{MN}\\&
+\int_{{\cal M}}d^{D}x \sqrt{g}\bigg[- \nabla^{M}\phi\nabla^{N}\phi^{*} + \frac{1}{2}g^{MN}\left(\vert\pa_{M}\phi\vert^2+\phi^*(m^2-\xi R)\phi\right)+ \xi |\phi|^2 R^{MN} \\&
+ \xi \left(g^{MN}\nabla_{K}\nabla^{K}- \nabla^{M}\nabla^{N}\right)|\phi|^{2}\bigg]\delta g_{MN}\\&
+\int_{\cal M}d^{D}x\sqrt{g}\left[-g^{MN}\nabla_{M}\nabla_{N}\phi^{*}+ (m^{2}-\xi R)\phi^{*}\right]\delta\phi \\&
+ \int_{{\cal M}}d^{D}x\sqrt{g}\left[-g^{MN}\nabla_{M}\nabla_{N}\phi+ (m^{2}-\xi R)\phi\right]\delta\phi^{*}\\&
+\int_{\pa {\cal M}}d^{D-1}x\sqrt{\beta}\bigg[\left(-\xi\nabla_{n}|\phi|^2 - \frac{1}{2}b|\phi|^2 -\frac{1}{2}a K |\phi|^{2}\right)\beta^{ij} \\& \quad\quad\quad\quad\quad\quad\quad\quad   
+ \left(\frac{1}{2\kappa} +\xi |\phi|^2\right){(K^{ij}- \frac{1}{D-1}K \beta^{ij})}
\bigg]\delta\beta_{ij} \\&
+\int_{\pa{\cal M}}d^{D-1}x |\phi|^{2}\left[2\xi\left(\frac{D-2}{D-1}\right)- a\right] \delta K \\&
+\int_{\pa{\cal M}}d^{D-1}x\sqrt{\beta}\left[\nabla_{n}\phi^{*} - b\phi^{*}- \frac{2}{D-1}K\xi\phi^{*}- aK \phi^{*}\right]\delta\phi \\&
+\int_{\pa{\cal M}}d^{D-1}x\sqrt{\beta}\left[\nabla_{n}\phi - b\phi - \frac{2}{D-1}K \xi \phi - aK \phi\right]\delta\phi^{*}\,.
}
The boundary terms of the above variation were presented in Eq.~\eqref{conformal_total_variation} of Section~\ref{subsec:Non-minimally coupled scalar + conformal boundary}.

\bibliographystyle{utphys}
\bibliography{Microcanonical_ref}

\providecommand{\href}[2]{#2}\begingroup\raggedright\begin{thebibliography}{10}

\bibitem{PhysRevD.33.2092}
J.~W. York, ``Black-hole thermodynamics and the euclidean einstein action,''
  \href{http://dx.doi.org/10.1103/PhysRevD.33.2092}{{\em Phys. Rev. D}
  {\bfseries 33} (Apr, 1986) 2092--2099}.
  \url{https://link.aps.org/doi/10.1103/PhysRevD.33.2092}.

\bibitem{PhysRevD.15.2752}
G.~W. Gibbons and S.~W. Hawking, ``Action integrals and partition functions in
  quantum gravity,'' \href{http://dx.doi.org/10.1103/PhysRevD.15.2752}{{\em
  Phys. Rev. D} {\bfseries 15} (May, 1977) 2752--2756}.
  \url{https://link.aps.org/doi/10.1103/PhysRevD.15.2752}.

\bibitem{PhysRevD.47.1420}
J.~D. Brown and J.~W. York, ``Microcanonical functional integral for the
  gravitational field,'' \href{http://dx.doi.org/10.1103/PhysRevD.47.1420}{{\em
  Phys. Rev. D} {\bfseries 47} (Feb, 1993) 1420--1431}.
  \url{https://link.aps.org/doi/10.1103/PhysRevD.47.1420}.

\bibitem{Draper_2022}
P.~Draper and S.~Farkas, ``Euclidean de sitter black holes and microcanonical
  equilibrium,'' \href{http://dx.doi.org/10.1103/physrevd.105.126021}{{\em
  Physical Review D} {\bfseries 105} no.~12, (June, 2022) }.
  \url{http://dx.doi.org/10.1103/PhysRevD.105.126021}.

\bibitem{Avramidi:1997sh}
I.~G. Avramidi and G.~Esposito, ``{Lack of strong ellipticity in Euclidean
  quantum gravity},'' \href{http://dx.doi.org/10.1088/0264-9381/15/5/006}{{\em
  Class. Quant. Grav.} {\bfseries 15} (1998) 1141--1152},
  \href{http://arxiv.org/abs/hep-th/9708163}{{\ttfamily arXiv:hep-th/9708163}}.

\bibitem{Anderson_2008}
M.~T. Anderson, ``On boundary value problems for einstein metrics,''
  \href{http://dx.doi.org/10.2140/gt.2008.12.2009}{{\em Geom. and Top.}
  {\bfseries 12} no.~4, (July, 2008) 2009--2045}.
  \url{http://dx.doi.org/10.2140/gt.2008.12.2009}.

\bibitem{Witten:2018lgb}
E.~Witten, ``{A note on boundary conditions in Euclidean gravity},''
  \href{http://dx.doi.org/10.1142/S0129055X21400043}{{\em Rev. Math. Phys.}
  {\bfseries 33} no.~10, (2021) 2140004},
  \href{http://arxiv.org/abs/1805.11559}{{\ttfamily arXiv:1805.11559
  [hep-th]}}.

\bibitem{Liu:2024ymn}
X.~Liu, J.~E. Santos, and T.~Wiseman, ``{New Well-Posed boundary conditions for
  semi-classical Euclidean gravity},''
  \href{http://dx.doi.org/10.1007/JHEP06(2024)044}{{\em JHEP} {\bfseries 06}
  (2024) 044}, \href{http://arxiv.org/abs/2402.04308}{{\ttfamily
  arXiv:2402.04308 [hep-th]}}.

\bibitem{Barvinsky_1996}
A.~Barvinsky and S.~Solodukhin, ``Non-minimal coupling, boundary terms and
  renormalization of the einstein-hilbert action and black hole entropy,''
  \href{http://dx.doi.org/10.1016/0550-3213(96)00438-5}{{\em Nuclear Physics B}
  {\bfseries 479} no.~1–2, (Nov., 1996) 305–318}.
  \url{http://dx.doi.org/10.1016/0550-3213(96)00438-5}.

\bibitem{Jacobson_2014}
T.~Jacobson and A.~Satz, ``On the renormalization of the gibbons-hawking
  boundary term,'' \href{http://dx.doi.org/10.1103/physrevd.89.064034}{{\em
  Physical Review D} {\bfseries 89} no.~6, (Mar., 2014) }.
  \url{http://dx.doi.org/10.1103/PhysRevD.89.064034}.

\bibitem{Neri_2023}
G.~Neri and S.~Liberati, ``On the resilience of the gravitational variational
  principle under renormalization,''
  \href{http://dx.doi.org/10.1007/jhep10(2023)054}{{\em Journal of High Energy
  Physics} {\bfseries 2023} no.~10, (Oct., 2023) }.
  \url{http://dx.doi.org/10.1007/JHEP10(2023)054}.

\bibitem{bastianelli2013oneloop}
F.~Bastianelli and R.~Bonezzi, ``One-loop quantum gravity from a worldline
  viewpoint,'' \href{http://arxiv.org/abs/1304.7135}{{\ttfamily arXiv:1304.7135
  [hep-th]}}.

\bibitem{Arnowitt:1962hi}
R.~L. Arnowitt, S.~Deser, and C.~W. Misner, ``{The Dynamics of general
  relativity},'' \href{http://dx.doi.org/10.1007/s10714-008-0661-1}{{\em Gen.
  Rel. Grav.} {\bfseries 40} (2008) 1997--2027},
  \href{http://arxiv.org/abs/gr-qc/0405109}{{\ttfamily arXiv:gr-qc/0405109}}.

\bibitem{PhysRevD.47.1407}
J.~D. Brown and J.~W. York, ``Quasilocal energy and conserved charges derived
  from the gravitational action,''
  \href{http://dx.doi.org/10.1103/PhysRevD.47.1407}{{\em Phys. Rev. D}
  {\bfseries 47} (Feb, 1993) 1407--1419}.
  \url{https://link.aps.org/doi/10.1103/PhysRevD.47.1407}.

\bibitem{Draper:2022xzl}
P.~Draper and S.~Farkas, ``{de Sitter black holes as constrained states in the
  Euclidean path integral},''
  \href{http://dx.doi.org/10.1103/PhysRevD.105.126022}{{\em Phys. Rev. D}
  {\bfseries 105} no.~12, (2022) 126022},
  \href{http://arxiv.org/abs/2203.02426}{{\ttfamily arXiv:2203.02426
  [hep-th]}}.

\bibitem{Draper:2023bhg}
P.~Draper, S.~Farkas, and M.~Karydas, ``{Path integral factorization and the
  gravitational effective action},''
  \href{http://dx.doi.org/10.1088/1361-6382/ad1449}{{\em Class. Quant. Grav.}
  {\bfseries 41} no.~2, (2024) 025004},
  \href{http://arxiv.org/abs/2310.02101}{{\ttfamily arXiv:2310.02101
  [hep-th]}}.

\bibitem{Vassilevich_2003}
D.~Vassilevich, ``Heat kernel expansion: user's manual,''
  \href{http://dx.doi.org/10.1016/j.physrep.2003.09.002}{{\em Physics Reports}
  {\bfseries 388} no.~5--6, (Dec., 2003) 279--360}.
  \url{http://dx.doi.org/10.1016/j.physrep.2003.09.002}.

\bibitem{Becker_2012}
D.~Becker and M.~Reuter, ``Running boundary actions, asymptotic safety, and
  black hole thermodynamics,''
  \href{http://dx.doi.org/10.1007/jhep07(2012)172}{{\em Journal of High Energy
  Physics} {\bfseries 2012} no.~7, (July, 2012) }.
  \url{http://dx.doi.org/10.1007/JHEP07(2012)172}.

\bibitem{DMMcAvity_1991}
D.~M. McAvity and H.~Osborn, ``Asymptotic expansion of the heat kernel for
  generalized boundary conditions,''
  \href{http://dx.doi.org/10.1088/0264-9381/8/8/010}{{\em Classical and Quantum
  Gravity} {\bfseries 8} no.~8, (Aug, 1991) 1445}.
  \url{https://dx.doi.org/10.1088/0264-9381/8/8/010}.

\bibitem{Dowker:1997mn}
J.~S. Dowker and K.~Kirsten, ``{Heat kernel coefficients for oblique boundary
  conditions},'' \href{http://dx.doi.org/10.1088/0264-9381/14/9/004}{{\em
  Class. Quant. Grav.} {\bfseries 14} (1997) L169--L175},
  \href{http://arxiv.org/abs/hep-th/9706129}{{\ttfamily arXiv:hep-th/9706129}}.

\bibitem{Elizalde_1999}
E.~Elizalde and D.~V. Vassilevich, ``Heat kernel coefficients for chern-simons
  boundary conditions in qed,''
  \href{http://dx.doi.org/10.1088/0264-9381/16/3/013}{{\em Classical and
  Quantum Gravity} {\bfseries 16} no.~3, (Jan., 1999) 813–822}.
  \url{http://dx.doi.org/10.1088/0264-9381/16/3/013}.

\bibitem{Avramidi:1997hy}
I.~G. Avramidi and G.~Esposito, ``{Gauge theories on manifolds with
  boundary},'' \href{http://dx.doi.org/10.1007/s002200050539}{{\em Commun.
  Math. Phys.} {\bfseries 200} (1999) 495--543},
  \href{http://arxiv.org/abs/hep-th/9710048}{{\ttfamily arXiv:hep-th/9710048}}.

\end{thebibliography}\endgroup
\end{document}